\documentstyle[12pt]{article}
\catcode`\@=11 % This allows us to modify PLAIN macros.
\def\maketitle{\par
\begingroup
\def\thefootnote{\fnsymbol{footnote}}
\def\@makefnmark{\hbox
to 0pt{$^{\@thefnmark}$\hss}}
\if@twocolumn
\twocolumn[\@maketitle]
\else
\global\@topnum\z@ \@maketitle \fi\thispagestyle{plain}\@thanks
\endgroup
\setcounter{footnote}{0}
\let\maketitle\relax
\let\@maketitle\relax
\gdef\@thanks{}\gdef\@author{}\gdef\@title{}\let\thanks\relax}
\def\@maketitle{
\null
\vskip 2em \begin{center}
{\LARGE \@title \par} \vskip 1.5em {\large \lineskip .5em
\begin{tabular}[t]{c}\@author
\end{tabular}\par}
\vskip 1em {\large \@date} \end{center}
\par
\vskip 1.5em}
\catcode`\@=12 % at signs are no longer letters
\newcommand {\pl}{\partial}

\newcommand {\vp}{\varphi}

\newcommand {\al}{\alpha}
\newcommand {\be}{\beta}
\newcommand {\ga}{\gamma}
\newcommand {\Ga}{\Gamma}

\newcommand {\la}{\lambda}
\newcommand {\La}{\Lambda}

\newcommand {\om}{\omega}

\newcommand {\ep}{\epsilon}

\newcommand {\del}  {\delta}
\newcommand {\Del}  {\Delta}

\newcommand {\half}{ {\frac{1}{2}} }

\newcommand {\sqg} {\sqrt{g}}

\newcommand {\Lcal}{{\cal L}}

\newcommand {\Dcal}{{\cal D}}

\newcommand {\Stil}{{\tilde S}}

\newcommand {\Zhat}{{\hat Z}}
\newcommand {\Gahat}{{\hat \Gamma}}
\newcommand {\altil}{{\tilde \alpha}}
\newcommand {\betil}{{\tilde \beta}}
\newcommand {\latil}{{\tilde \lambda}}
\newcommand {\Ctil}{{\tilde C}}
\newcommand {\labar}{{\bar \lambda}}

\newcommand {\PLinv}  {{\frac{1}{\hbar}}}
\newcommand {\intx} {{\int d^2x}}

\newcommand {\expect}  {A<\int d^2x\sqrt{g} R^2>}
\newcommand {\change} {\leftrightarrow}
\newcommand {\ra} {\rightarrow}
\newcommand {\pr}   {{\quad .}}
\newcommand {\com}  {{\quad ,}}
\newcommand {\q}    {\quad}
\newcommand {\qq}   {\quad\quad}
\newcommand {\qqq}   {\quad\quad\quad}
\newcommand {\qqqq}   {\quad\quad\quad\quad}

\newcommand {\nn}    {\nonumber}
\newcommand {\vs}[1]  { \vspace*{#1 cm} }
\newcounter{eq}
\newcounter{sc}

\newcommand {\MPL}  {Mod.Phys.Lett.}
\newcommand {\NP}   {Nucl.Phys.}
\newcommand {\PL}   {Phys.Lett.}
\newcommand {\PR}   {Phys.Rev.}

\begin{document}

\begin{flushright}
US-95-06\\
\end{flushright}
\vspace{-1.0cm}

%\renewcommand{\baselinestretch}[2]
%\begin{document}
%\begin{flushright}
%RIMS-999
%\end{flushright}
\title{    Thermodynamic Properties,Phases and
           Classical Vacua of Two Dimensional
           $R^2$-Gravity
                                            }
%\title{    Classical Vacua of Two Dimensional
%           $R^2$-Gravity and Thermodynamic Properties
%                                            }
\author{ Shoichi ICHINOSE
         \thanks{ E-mail address:ichinose@u-shizuoka-ken.ac.jp}   \\
         Department of Physics, Universuty of Shizuoka,           \\
         Yada 52-1, Shizuoka 422, Japan
                }
\date{  July, 1995 }
\maketitle
\setlength{\baselineskip}{0.8cm}
\begin{abstract}
Two dimensional quantum R$^2$-gravity is formulated
in the semiclassical method.
The thermodynamic properties,such as
the equation of state,
the temperature and the entropy, are explained.
The topology constraint and the area constraint are properly taken
into account. A total derivative term
and an infrared regularization play  important roles.
The classical solutions (vacua)
of R$^2$-Liouville equation are obtained by making use of the well-known
solution of the
ordinary Liouville equation.
The positive and negative constant
curvature solutions are 'dual' each other.
Each solution has two branches($\pm$).
We characterize all phases.
The topology of a sphere is mainly considered.

\end{abstract}
%%%%%%%%%%%%%%%%%%%%%%%%%%%%%%%%%%%%%%%%%%
%%%%%%%%%%%%%%%%%%%%%%%%%%%%%%%%%%%%%%%%%%%%%%%%%%%%%%%%%%%%%%%%%%%%%
%%%%%%%%%%%%%%%%%%%%%%%%%%%%%%   SEC  1    %%%%%%%%%%%%%%%%%%%%%%%%%%
%%%%%%%%%%%%%%%%%%%%%%%%%%%%%%%%%%%%%%%%%%%%%%%%%%%%%%%%%%%%%%%%%%%%%
\section{Introduction}

$\qq$ In the recent analysis of two dimensional(2d) quantum gravity(QG), the
conformal approach or the matrix model approach have been intensively
done. Those approaches are nonperturbative ones and
it is expected that some non-perturbative features are
important to understand the theory. At the same time, however,
it is known that an orthodox perturbative approach, the semiclassical
approach, is also useful in 2d QG\cite{DJ,SEI}.
We present a close examination  of the latter approach.
A key point in this treatment is how to treat the area constraint
and the topology constraint. The regularization of infrared divergence
(and ultraviolet divergence in the quantum evaluation) is also important.
The semiclassical treatments of 2d QG so far are insufficient in these
points. We present a new formalism.

Despite of the long period of research, the physical picture of the
Liouville theory, in relation to 2d QG, seems obscure.
It shows the delicacy or the subtlety
of the theory and requires some other proper
formalism and regularization.
As far as the popular formalism based on the conformal field theory
is taken,
the barrier $c_m=1$~ does not seem to be overcome.
Whereas the computer simulations seem to no special difficulty for
the  prohibitted region of $c_m$\cite{AJT,AT}.
This conflict seems gradually serious.
Although the problem is examined from different approaches,
it is fair to say the true situation is
not known at present. Recently it has been shown that
the semiclassical results nicely explain
the simulation data\cite{ITY1,ITY2}.

It is well known, in the lattice QG,
that a higher-derivative term, $\be R^2$, regularize the theory very well.
It has the effect of smoothing  the surface (if we take
a proper coupling sign).
The importance of the term is also stressed in the continuum context\cite{KPZ}.
{}From the simple power-counting we  see the ultra-violet
behaviour becomes well regularized.
We can take two standpoints about the $R^2$-term:\
1)\ We are considering the  2d $R^2$-QG as one gravitational model;\
2)\ We regard $R^2$-term as a regularization to define the $\be=0$~theory
and expect its effect
disappears when some limit is taken.
Although 1) is mainly taken in the present paper,
both standpoints are important at this time of development.

In the semiclassical analysis of 2d $R^2$-QG in \cite{ITY1,ITY2}
one of
the classical solutions (positive curvature solution)
is analysed. In the present paper we present
the full structure of the classical vacua.
It is quite interesting that the positive and negative solutions are
symmetric with respect to a reflection in the coupling $\be$-space.
The explanation is self-contained.

It is well known that the global quantities in the gravitational
system, such as entropy, volume and temperature of the total universe, obey
the laws of thermodynamics. It says those quantites can be
regarded as thermodynamic state variables of an equilibrium state. We will find
those properties in the present 2d model of QG. We can characterize all phases
appearing in the theory by the $\be$-dependence of the temperature.

We present a general formalism in Sec.2, where some thermodynamic
functions are introduced. In Sec.3 the classical solutions are obtained.
They are $R^2$-gravity
version of the Liouville solution and describe positive
and negative constant-curvature manifolds. The analytic expressions
of some physical quantites are given and analysed.
We characterize all asymptotic regions of the solutions in Sec.4.
In Sec.5 an integral about
a parameter $\la$, which appears in the general formalism in connection with
the area costraint, is done. The analytic expressions of cross-over
points in the theory are obtained in Sec.6. It shows the essetial
behaviour of the theory is controled by a toatl derivative (global) term.
In Sec.7 we characterize each phase and obtain the equation of state.
The expressions for temperature and entropy are obtained.
We conclude in Sec.8.
%%%%%%%%%%%%%%%%%%%%%%%%%%%%%%%%%%%%%%%%%%%%%%%%%%%%%%%%%%%%%%%%%%%%%
%%%%%%%%%%%%%%%%%%%%%%%%%%%%%%   SEC  2    %%%%%%%%%%%%%%%%%%%%%%%%%%
%%%%%%%%%%%%%%%%%%%%%%%%%%%%%%%%%%%%%%%%%%%%%%%%%%%%%%%%%%%%%%%%%%%%%
\section{Semiclassical Quantization and Thermodynamic Functions}

$\qq$ Before the concrete evaluation, we describe here the present new
formalism.
We take the Euclidean action,
%****(3.0)%%%%%%%%%%%%%%%%%%%%%
\begin{eqnarray}
& S_{tot}=S_{gra}+S_m\com \q
 S_{gra}[g;G,\be,\mu]=\intx\sqg (\frac{1}{G} R-\be R^2-\mu)\com
                                                              & \nn\\
& S_m[g,\Phi;c_m]=-\intx\sqg (\half\sum_{i=1}^{c_m}\pl_a\Phi_i\cdot
g^{ab}\cdot \pl_b\Phi_i)\com\q (\ a,b=1,2\ )\com            & \label{3.0}
\end{eqnarray}
%%%%%
under the fixed area condition
$ A=\intx \sqg\ $. Here
$G$\ is the gravitaional coupling constant, $\mu$\ is the cosmological
constant ,
$\be$\ is the coupling strength for $R^2$-term and $\Phi$\ is the $c_m$-
components scalar matter fields.

The 2 dim quantum gravity can be treated in the way similar to
the flat theory by
taking the conformal-flat gauge($a,b=1,2$),
%***(3.1)%%%%%%%%%%%%%%%%%%%%%
\begin{eqnarray}
& g_{ab}=\ e^{\vp}\ \del_{ab}\com
  (\del_{ab})=
               \left( \begin{array}{cc} 1 & 0 \\
                                        0 & 1
                      \end{array}\right)
                                         \com&            \label{3.1}
\end{eqnarray}
%%%%%%%%%%%%%%%%%%%%%%%%%%%%%
the action~(\ref{3.0})
gives us,after integrating out the matter fields and Faddeev-Popov ghost,
the following partition function\cite{P}.
%****(3.3) (3.2)%%%%%%%%%%%%%%%%%%%%
\begin{eqnarray}
& \int\frac{\Dcal g\Dcal\Phi}{V_{GC}}\{exp\PLinv S_{tot}\}~\del(\intx\sqg-A)
=exp\PLinv (\frac{8\pi(1-h)}{G}-\mu A)\times Z[A]\com & \nn\\
& Z[A]\equiv\int\Dcal\vp~ e^{+\frac{1}{\hbar}
S_0[\vp]}~\del(\intx ~e^\vp - A)\com &   \label{3.3}\\
& S_0[\vp]=\intx\ (\frac{1}{2\ga}\vp\pl^2\vp
-\be~e^{-\vp}(\pl^2\vp)^2 +\frac{\xi}{2\ga}\pl_a(\vp\pl_a\vp)\ )\com
\q \frac{1}{\ga}=\frac{1}{48\pi}(26-c_m)\com & \label{3.2}
%\pl^2=(\frac{\pl}{\pl x})^2+(\frac{\pl}{\pl y})^2\com
\end{eqnarray}
%%%%%%%%%%%%%%%%%%%%%%%%%%%%%
where  the  relations for
Einstein term and the cosmological term:\
$\intx\sqg R =8\pi (1-h),\ h=\mbox{number of handles},
\intx\sqg =A $\ ,are used\cite{N2a}.
%\footnote{
%The sign for the action is different from the usual convention as seen in
%(\ref{3.3}).
%}
$V_{GC}$\ is the gauge volume due to the general coordinate invariance.
$\xi$\ is a free parameter. The total derivative term generally appears when
integrating out the anomaly equation
\ $\del S_{ind}[\vp]/\del\vp=\frac{1}{\ga}\pl^2\vp\ $.
This term turns out to be very important\cite{N2b}.
%\footnote{
%The uniqueness of this term, among all possible total derivatives, is shown
%in Discussions(sect.6) of \cite{ITY1}.
%}
We consider the manifold of a fixed topology of
the sphere ,$h=0$, and with the finite area $A$.
Furthermore we consider the case $\ga>0\ (c_m<26)$\cite{N2c}.
%\footnote{
%This is for the comparison with the 'classical limit' $c_m\ra -\infty$.
%We can do the same analysis for $\ga<0$\ without any difficulty.
%}
%The case of $c_m=0\ (\ga=\frac{24}{13}\pi)$ corresponds to the pure
%(R$^2$-) gravity.
\ $\hbar$\ is  Planck constant\cite{N2d}.
%%%%% foot
%\footnote{
%In this section only,we explicitly write $\hbar$\ (Planck constant) in order
%to show the perturbation structure clearly.
%}

The Laplace transform of (\ref{3.3}) is given by
%***(3.4a)%%%%%%%%%%%%%%%%%%%%%
\begin{eqnarray}
&{\Zhat}[\la]=
\int_0^{\infty} Z[A]e^{-\la A/\hbar}~dA   &  \nn  \\
&\qqq=\int\Dcal\vp~exp[~
+\frac{1}{\hbar}\{ S_0[\vp]-\la\ \intx ~e^\vp \} ]\pr &
                                                      \label{3.4a}
\end{eqnarray}
%%%%%%%%%%%%%%%%%%%%%%%%%%%%%
As the arguments of
 $Z$,(\ref{3.3}), and $\Zhat$,(\ref{3.4a}),
we do not write explicitly $\be,\ga$-
dependence .
(\ref{3.3}) is the micro canonical distribution for the fixed area $A$\ ,
whereas (\ref{3.4a}) is the grand canonical distribution (variable
area  ) with the chemical potential $\la$\ . From (\ref{3.4a}), we obtain
the expectation value for the area as
%****(3.4c)%%%%%%%%%%%%%%%%%%%%
\begin{eqnarray}
<A_{op}>=\frac{1}{\Zhat}\frac{d}{d(-\la/\hbar )}\Zhat[\la]
\equiv <\intx e^{\vp}>_{\Zhat}\com\q
A_{op}\equiv\intx~e^\vp\pr                  \label{3.4c}
\end{eqnarray}
%%%%%%%%%%%%%%%%%%%%%%%%%%%%%
By inverting this equation we obtain \ $\la=\labar(<A_{op}>)$\ . Equivalently
we also abtain it in terms of the Legendre-transformed generating
function $\Ga[<A_{op}>]$.
%****(3.4d)%%%%%%%%%%%%%%%%%%
\begin{eqnarray}
\Ga[<A_{op}>]\equiv ln~\Zhat[\labar]+\PLinv\labar\times <A_{op}>\com\ \
\labar(<A_{op}>)=\hbar\frac{d\Ga[<A_{op}>]}{d<A_{op}>}\pr
                                                    \label{3.4d}
\end{eqnarray}
%%%%%%%%%%%%%%%%%%%%%%%%%%%%%

$Z[A]$\ can be obtained from $\Zhat[\la]$\ by the inverse Laplace
transformation.
%***(3.4f)%%%%%%%%%%%%%%%%%%
\begin{eqnarray}
& Z[A]=\int d\la~\Zhat[\la]~e^{+\la A/\hbar}
\equiv\int d\la~ Y[A,\la]\com & \nn\\
& Y[A,\la]\equiv
\int\Dcal\vp~exp\ \frac{1}{\hbar}[\ S_0[\vp]
-\la (\intx e^\vp - A)]      &   \label{3.4f}\\
& \Ga^{eff}[A,\la]\equiv~ln~Y[A,\la]\com\nn
\end{eqnarray}
%%%%%%%%%%%%%%%%%%%%%%%%%%%%%
where $\la$-integral should be carried out along an appropriate contour
parallel to the imaginary axis of the complex $\la$-plane(Fig.1).

{\vs 5}
\begin{center}
Fig.1\q The contour of $\la$-integral in the complex $\la$-plane
\end{center}

$\Zhat[\la]$ is defined
by (\ref{3.4a}). The partition function (\ref{3.4f}) can be evaluated
semiclassically in the following two steps (i) and (ii).
In the evaluation we will
relate above thermodynamic functions.
%%%%%%%%%%%%%%%%%%%%%%% (i) \vp-integral  %%%%%%%%%%%%%%%%%%%%%%%%%
\flushleft{(i)\ $\vp$-integral}

First we define some quantities.
%***(3.4g)%%%%%%%%%%%%%%%%%%%%
\begin{eqnarray}
& S_\la[\vp]\equiv S_0[\vp]-\la\intx~e^\vp\  & \nn\\
& =\intx\ (\frac{1}{2\ga}\vp\pl^2\vp -\be~e^{-\vp}(\pl^2\vp)^2\
             +\frac{\xi}{2\ga}\pl_a(\vp\pl_a\vp) -\la~e^\vp\ )\com    &\nn\\
&\Zhat[\la]=\int\Dcal\vp~exp~\{\PLinv S_\la[\vp]\}
\equiv\ exp~\PLinv\Gahat[\la]&
                                                      \label{3.4g}
\end{eqnarray}
%%%%%%%%%%%%%%%%%%%%%%%%%%%%%
$\Gahat(\la)$\ is the effective action corresponding to $S_\la[\vp]$.
$\Gahat(\la)$ can be evaluated loop-wise\cite{N2e}
%\footnote{
%The expansion parameter is the Planck constant $\hbar$\ . It is known,
%in the field theory, the expansion with repect to $\hbar$\ is equivalent
%to that with respect to the number of loops in the Feynman diagrams
%(loop-expansion).
%}
by the semiclassical expansion.
%***(3.4h)%%%%%%%%%%%%%%%%%%%
\begin{eqnarray}
\vp(x)=\vp_c(x;\la)+\sqrt{\hbar}~\psi(x)\com
                                       \label{3.4h}
\end{eqnarray}
%%%%%%%%%%%%%%%%%%%%%%%%%%%%%
where we take the 'mean field' (or 'background field') as the solution
of the classical field equation for $S_\la[\vp]$.
%***(3.4i)%%%%%%%%%%%%%%%%%%%
\begin{eqnarray}
\left.\frac{\del}{\del\vp}S_\la[\vp]\right|_{\vp_c}=\ 0\pr
                                       \label{3.4i}
\end{eqnarray}
%%%%%%%%%%%%%%%%%%%%%%%%%%%%%
Then $\Gahat[\la]=\hbar~ln~\Zhat[\la;\vp]$ can be evaluated as
%***(3.4j)%%%%%%%%%%%%%%%%%%%
\begin{eqnarray}
&\Zhat[\la]=\int \Dcal\psi~
          exp~\PLinv S_\la[\vp_c+\sqrt{\hbar}\psi]      &\nn\\
&=~exp~\PLinv S_\la[\vp_c]\times \int \Dcal\psi~
exp\{~\frac{1}{2}\frac{\del^2S_\la}{\del\vp^2}|_{\vp_c}\psi\psi
                   +O(\sqrt{\hbar}\psi^3) \}    &\label{3.4j}\\
&\equiv exp\{~\PLinv\Gahat^0[\la]+\Gahat^1[\la]+
O(\mbox{higher-than 1-loop,\ }\hbar )~\}\com    & \nn\\
&\Gahat^0[\la]\equiv S_\la[\vp_c]\com&           \nn\\
&\Gahat[\la]=\Gahat^0[\la]+\hbar\Gahat^1[\la]
+O(\mbox{higher-than 1-loop,\ }\hbar^2 )\com&\nn
\end{eqnarray}
%%%%%%%%%%%%%%%%%%%%%%%%%%%%%%%%
where $\Gahat^n[\la],(n\geq 1),$\ is the n-loop quantum effects.
In (\ref{3.4j}),
the effect up to 1-loop order is explicitly written.
%%%%%%%%%%%%%%%%%%%%%%% (ii) \vp-integral  %%%%%%%%%%%%%%%%%%%%%%%%%
\flushleft{(ii)\ $\la$-integral}

Using the result of (i), $Z[A]$ can be written as
%***(3.4k)%%%%%%%%%%%%%%%%%%%%
\begin{eqnarray}
Z[A]=\int d\la~exp~\PLinv\{~\Gahat[\la]+\la A~\}\pr
                                       \label{3.4k}
\end{eqnarray}
%%%%%%%%%%%%%%%%%%%%%%%%%%%%%
The $\la$-integral of (\ref{3.4k}) can be again evaluated in the
semiclassical way as follows. The dominant value $\la_c$\ is defined by,
%***(3.4l)%%%%%%%%%%%%%%%%%%%%
\begin{eqnarray}
\frac{d}{d\la}(\Gahat[\la]+\la A)|_{\la_c}
=\frac{d\Gahat[\la_c]}{d\la_c}+A=0\com \nn\\
\Gahat[\la]=\Gahat^0[\la]+\hbar\Gahat^1[\la]+\cdots\com\q
\la_c=\la_c^0+\hbar\la_c^1+\cdots\com\label{3.4l}
\end{eqnarray}
%%%%%%%%%%%%%%%%%%%%%%%%%%%%%
where $\la_c^n$ \ is the n-loop effect of $\psi$-integration and
is recursively obtained.
The $\la$-integral is approximately obtained by
evaluating the fluctuation around $\la_c$\  perturbatively
as follows.
%***(3.4ll)%%%%%%%%%%%%%%%%%%%%
\begin{eqnarray}
& \la=\la_c+~\om\com                            &       \nn\\
& Z[A]
=exp~\PLinv\{\Gahat[\la_c]+\la_c A\}\times\int d\om~
exp~\PLinv\{~\frac{1}{2}
\left.\frac{d^2\Gahat[\la]}{d\la^2}\right|_{\la_c}~\om^2
                        +O(\om^3) \ \}\pr      &\label{3.4ll}
\end{eqnarray}
%%%%%%%%%%%%%%%%%%%%%%%%%%%%%
This integral will be evaluated in Sec.5.

$\qq$ Here we note that the equation (\ref{3.4l}),
by which $\la_c$\ is defined,
is re-written as
%***(3.4m)%%%%%%%%%%%%%%%%%%%%
\begin{eqnarray}
A=\ -\frac{d\Gahat[\la_c]}{d\la_c}=\left.\frac{1}{\Zhat[\la]}
\frac{d\Zhat[\la]}{d(-\la/\hbar)}\right|_{\la_c}\com
                                        \label{3.4m}
\end{eqnarray}
%%%%%%%%%%%%%%%%%%%%%%%%%%%%%
which is exactly the same as (\ref{3.4c})
by identifying $A$\ above with $<A_{op}>$ \ in (\ref{3.4c}).
Therefore $\la_c(A)=\labar(A)$\ . And
$exp\{\PLinv\Gahat[\la_c]\}=~\Zhat[\la_c]$ ,in
the second equation of(\ref{3.4ll}) ,is exactly
the same quantity as $\Zhat[\labar(A)]$ in  (\ref{3.4d}).
Furthermore $\Ga^{eff}[A,\la_c]$\
exactly coincides
with $\Ga[A]$\ of (\ref{3.4d}).
%***(3.4n)%%%%%%%%%%%%%%%%%%%%
\begin{eqnarray}
ln~Z[A]\approx\Ga^{eff}[A,\la_c]=\PLinv(\Gahat[\la_c]+\la_c A\ )
                                          =\Ga[A]\pr  \label{3.4n}
\end{eqnarray}
%%%%%%%%%%%%%%%%%%%%%%%%%%%%%

$\qq$ We have introduced some thermodynamic functions. For their comparison
they are listed in Table 1.

\vspace{0.5cm}
%%%%%%%%%%%%%%%%%%%%%%%%%%%%%%%%%%%%%%%%%%%%%%%%%%%%%%%%%%%%%%%%%%%%%%%%%%
%%%%%%%%%%%%%%%%%%%%%  Table 1   %%%%%%%%%%%%%%%%%%%%%%%%%%%%%%%%%%%%%%%%%
%%%%%%%%%%%%%%%%%%%%%%%%%%%%%%%%%%%%%%%%%%%%%%%%%%%%%%%%%%%%%%%%%%%%%%%%%%
\begin{tabular}{|c|c|c|c|c|}
\hline
Indep.    & Integral  & Special          & Partition Func.
                                                     & Effective      \\
param. & var.     &  func.           &               & action        \\
\hline
$A$       & $\vp(x)$     &            &  $ Z[A] $, (\ref{3.3})   &      \\
\hline
$\la$     & $\vp(x)$     & $<A_{op}>$, (\ref{3.4c}) &
                        $ \Zhat[\la] $, (\ref{3.4a})      &              \\
:real     &              & $\labar(<A_{op}>)$, (\ref{3.4c}) &        &
                                    $\Gahat[\la]=\hbar~ln~\Zhat[\la] $,  \\
          &              & $\Zhat[\labar(A)]$  &         &  (\ref{3.4g}) \\
\hline
$<A_{op}>$ &       &                    &        &
                                                 $ \Ga[<A_{op}>]$ ,      \\
           &             & $\labar(<A_{op}>)$,(\ref{3.4d}) &        &
                                                            (\ref{3.4d}) \\
\hline
$A$        & $\vp(x)$,    & $\vp_c(x,\la)$,(\ref{3.4i})      &
             $Y[A,\la]$,(\ref{3.4f}) &               \\
           & $\la$       & $\la_c(A)=\labar(A)$     &   &  $\Ga^{eff}[A,\la]$
 \\
           &:complex      & :real,(\ref{3.4l})  &
                                             & $=ln~Y[A,\la]$,(\ref{3.4f})   \\
           &             & $\Zhat[\la_c(A)]$   &             &               \\
           &             & $\Ga^{eff}[A,\la_c]$     &     &        \\
           &             & $=\Ga[A]$,(\ref{3.4n}) &    &        \\
\hline
\multicolumn{5}{c}{\q}                                   \\
\multicolumn{5}{c}{Table 1\q Some Thermodynamic Functions.}
\end{tabular}
%%%%%%%%%%%%%%%%%%%%%%%%%  END  of  Table 1 %%%%%%%%%%%%%%%%%%%%%%%%%%%%%
%%%%%%%%%%%%%%%%%%%%%%%%%%%%%%%%%%%%%%%%%%%%%%%%%%%%%%%%%%%%%%%%%%%%%%%%%
\vspace{0.5cm}
%%%%%%%%%%

$\qq$ We notice the area constraint ,which was expressed as the delta function
in (\ref{3.3}), is replaced by the $\la$-integral in the present
formalism using $Y[A,\la]$. This is the key point for
correctly treating the area constraint in the semiclassical method\cite{N2f}.
%\footnote{
%This method is so popular that its naming is diverse depending on the
%applied circumstances: \ mean-field method, WKB-approximation
%,stationary phase approximation, effective action method,
%background-field method,etc.
%          }.
The quantum effect is systematically evaluated loop-wise\cite{S1}:\ the
renormalization of parameters involved in the theory, due to the quantum
interaction of the Weyl mode $\vp$, is done in
eq.(\ref{3.4j}).

$\qq$ In the following,
we will evaluate the leading order,i.e. order of $\hbar^0$.
%%%%%%%%%%%%%%%%%%%%%%%%%%%%%%%%%%%%%%%%%%%%%%%%%%%%%%%%%%%%%%%%%%%%%
%%%%%%%%%%%%%%%%%%%%%%%%%%%%%%   SEC  4    %%%%%%%%%%%%%%%%%%%%%%%%%%
%%%%%%%%%%%%%%%%%%%%%%%%%%%%%%%%%%%%%%%%%%%%%%%%%%%%%%%%%%%%%%%%%%%%%
\section{Classical Vacua of R$^2$-Gravity}

$\qq$ The classical configuration (solution) of
the special case
,$\be=0$,  was well-known as the Liouville solutions.
(See \cite{SEI} for a recent review.)  Furthermore,
in the context of 2 dim quantum gravity (or the string theory ) ,
the special case was
already examined by \cite{OV} and \cite{Z} .
We consider the general case:\ $\be$\ is
the arbitray real number with the dimension of (Length)$^2$.

$\qq$ The classical field equation (\ref{3.4i}) is given by
%**** 3.6a    %%%%%%%%%%%%%%%%
\begin{eqnarray}
\frac{\del S_\la[\vp]}{\del\vp}=\frac{1}{\ga}\pl^2\vp
+\be\{ e^{-\vp}(\pl^2\vp)^2-2\pl^2(e^{-\vp}\pl^2\vp)\}-\la e^\vp=0\pr
                                                         \label{3.6a}
\end{eqnarray}
%%%%%%%%%%%%%%%%%%%%%%%%%%%%%
We make the following assumption of constant-curvature
for the  solution.
%*** 3.8    %%%%%%%%%%%%%%%%
\begin{eqnarray}
-R|_{\vp_c}=\
e^{-\vp_c}\pl^2\vp_c
=\mbox{const}\equiv \frac{-\al}{A}\com        \label{3.8}
\end{eqnarray}
%%%%%%%%%%%%%%%%%%%%%%%%%%%%%
where $\al$ is a dimensionless constant.
The above equation is the Liouville equation.
It is easy to find that the
solution (\ref{3.8}) satisfies (\ref{3.6a}) for
such real $\al$ that satisfies the following equation:
%*** 3.9a    %%%%%%%%%%%%%%%%
\begin{eqnarray}
\mbox{COND.1}\qqqq\al^2\be'-\frac{1}{\ga}\al-\la A=0\com\q
\be'\equiv\frac{\be}{A}\pr
                                               \label{3.9a}
\end{eqnarray}
%%%%%%%%%%%%%%%%%%%%%%%%%%%%%
We may safely assume the spherical symmetry (in (x,y)-plane)
for a stable solution:
$\vp_c=\vp_c(r),r=\sqrt{x^2+y^2}$. Then (\ref{3.8}) reduces to
%***  3.7   %%%%%%%%%%%%%%%%
\begin{eqnarray}
\frac{1}{r}\frac{d}{dr}r\frac{d\vp_c}{dr}+\frac{\al}{A}e^{\vp_c}=0
\pr                   \label{3.7}
\end{eqnarray}
%%%%%%%%%%%%%%%%%%%%%%%%%%%%%

$\qq$Before further anlysis , we comment on the
eq.(\ref{3.8}). The Liouville equation
(\ref{3.8}) corresponds to the ordinary gravity ($\be=0$)
\ :\ $S=\intx (\frac{1}{2\ga}\vp\pl^2\vp-\mu_1~e^\vp)$\ ,
with the
cosmological constant $\mu_1=-\frac{1}{\ga}\frac{\al}{A}$,
which is negative for $\al>0$ and positive for $\al<0$.
%%%%%%%%%%%%%%%%%%%%%%%%%%%%%%%%%%%%%%%%%%%%%%%%%%%%%%%%%%%
%%%%%%%%%%%%%%%%%%%%%%%%   4.1       %%%%%%%%%%%%%%%%%%%%%%
\subsection{Positive Curvature Solution}

$\qq$ For the case:
%*** 3.10b    %%%%%%%%%%%%%%%%
\begin{eqnarray}
\al >0\com                                            \label{3.10b}
\end{eqnarray}
%%%%%%%%%%%%%%%%%%%%%%%%%%%%%
the solution of (\ref{3.7}) is given by (cf.\cite{OV,Z,SEI}),
%*** 3.10a    %%%%%%%%%%%%%%%%
\begin{eqnarray}
\vp_c(r;\al )=-ln~\{ \frac{\al}{8}(1+\frac{r^2}{A})^2\}\pr  \label{3.10a}
\end{eqnarray}
%%%%%%%%%%%%%%%%%%%%%%%%%%%%%
This solution satisfies
$\left.\intx\sqg\right|_{\vp_c}=\intx~e^{\vp_c}=\frac{8\pi}{\al}A$\ ,
$-\left.\intx\sqg R\right|_{\vp_c}=\intx~\pl^2\vp_c=-8\pi$\ .
It means the manifold described by (\ref{3.10a})
has the area $\frac{8\pi}{\al}A$\ and
is topologically the sphere.
The equations (\ref{3.9a}
-\ref{3.10a}) constitute a solution of (\ref{3.6a}).

$\qq$ $S_{\la}[\vp_c]$ is given as
%*** 3.11    %%%%%%%%%%%%%%%%
\begin{eqnarray}
& S_{\la}[\vp_c]\left(=\Gahat^0(\la)\right)
=(1+\xi)\frac{4\pi}{\ga}~ln\frac{\al}{8}-16\pi\al\be'+C(A)\com & \nn\\
& C(A)=\frac{8\pi (2+\xi)}{\ga}+\frac{8\pi\xi}{\ga}
\{~ln(1+L^2/A)-(L^2/A)/(1+(L^2/A))~\}      &
                                             \label{3.11}\\
& =\frac{8\pi (2+\xi)}{\ga}+\frac{8\pi\xi}{\ga}
\{~ln~\frac{L^2}{A}-1~\}+O(\frac{A}{L^2})\com\q \frac{L^2}{A}\gg 1\com
                                                   &\nn
\end{eqnarray}
%%%%%%%%%%%%%%%%%%%%%%%%%%%%%
where
$L$\ is the infrared cut-off ($r^2\leq L^2$)
introduced for the divergent volume intgral of the total derivative term
($\xi$-term). See Fig.2.
The integral is log-divegent at $r \rightarrow\infty$\ for
the classical solution (\ref{3.10a}).
$\al$ (or $\la$)is rewritten in terms of $\la$ (or $\al$)
by use of (\ref{3.9a}) and $C(A)$\ does not depend on $\al$\ and $\be$\
but depends on $A$\ and $\ga$.
In the above derivation, formula in App.A are usefull.

{\vs 5}
\begin{center}
Fig.2\q Infra-red cut-off $L$\ in
the flat coordinates and the sphere manifold.
For simplicity, the picture is for $\al=8.$\ For general $\al$,
$(x,y,r,\sqrt{A},L)$\ is substituted by
$\sqrt{8/\al} \times (x,y,r,\sqrt{A},L)$.
\end{center}

$\qq$ The eq. (\ref{3.4l}) at the classical level is written as,
%***   3.16    %%%%%%%%%%%%%%%%
\begin{eqnarray}
\frac{dS_{\la}[\vp_c]}{d\la}+A
=(\frac{4\pi}{\ga}\frac{1}{\al}(1+\xi)
-16\pi\be')\frac{d\al}{d\la}+A                            \nn\\
=\{ \frac{4\pi}{\ga}\frac{1}{\al}(1+\xi)-(16\pi\be'+\frac{1}{\ga})
+2\be'\al\}\frac{d\al}{d\la}=0\com                   \label{3.16}
\end{eqnarray}
%%%%%%%%%%%%%%%%%%%%%%%%%%%%%
where we have used a relation :\ $1=\frac{d\la}{d\al}\frac{d\al}{d\la}
=\frac{1}{A}(~2\al\be'-\frac{1}{\ga})\frac{d\al}{d\la}\ $
, which is derived
from (\ref{3.9a}). For the case $\frac{d\al}{d\la}\neq 0$~\cite{N3a}
%\footnote{
%This condition turns out to be satisfied in the following solution
%          }
,(\ref{3.16}) says
%*** 3.17    %%%%%%%%%%%%%%%%
\begin{eqnarray}
\mbox{COND.2} & \Xi(\al;\be',\ga,\xi)\equiv
                2\be'\al^2-(16\pi\be'+\frac{1}{\ga})\al+
(1+\xi)\frac{4\pi}{\ga}=0\com &                          \nn\\
& \al^{\pm}_c=\frac{1}{4\be'}\{ 16\pi\be'+\frac{1}{\ga}
\pm\sqrt{D} ~\}                   \com &        \label{3.17}\\
& D\equiv (16\pi\be')^2
+\frac{1}{\ga^2}-\xi\frac{32\pi\be'}{\ga}
=(16\pi\be'-\frac{\xi}{\ga})^2+\frac{1-\xi^2}{\ga^2}\pr &    \nn
\end{eqnarray}
%%%%%%%%%%%%%%%%%%%%%%%%%%%%%
The relation (\ref{3.9a}) gives $\la^{\pm}_c(\be)
\equiv \la(\be,\al^{\pm}_c(\be))$. Note that the determinant of the above
quadratic equation ,$D$, is positive definite for all real
$\be$ if the following
condition is satisfied.
%***   3.17b    %%%%%%%%%%%%%%%%
\begin{eqnarray}
-1\leq\ \xi\ \leq\ +1\pr                   \label{3.17b}
\end{eqnarray}
%%%%%%%%%%%%%%%%%%%%%%%%%%%%%
We consider this case in the following.

$\qq$ In summary, 2 unknown variables $\al$\ and $\la$\ ,are fixed by
two conditions COND.1 and 2 and they are expreesed by three physical
parameters $\be$\ ,$\ga$\ ,$A$\ and  one free parameter $\xi$.
We list here the obtained result of important physical quantites.
%*** 3.18    %%%%%%%%%%%%%%%%
\begin{eqnarray}
\mbox{Curvature}   &
\al^{\pm}_c=\frac{4\pi}{w}\{ w+1~\pm\sqrt{w^2+1 -2\xi w} ~\}
                                                        \com\nn\\
\mbox{Classical Action} &
S_{\la}[\vp_c]
=(1+\xi)\frac{4\pi}{\ga}~ln~\frac{\al_c}{8}-\frac{w}{\ga}\al_c
                                        +C(A)    \com   \nn\\
\mbox{String Tension}  &
\la_c A=\frac{1}{16\pi\ga}({\al_c}^2w-16\pi\al_c)\com      \label{3.18}\\
\mbox{An Expect. Value} &
-A<\intx\sqg R^2>|_{c}=
\frac{\pl \Ga^{eff}[\vp_c]}{\pl\be'}                  \nn\\
&\qq =-16\pi\al_c+\al_c^2+\Xi\times
\frac{1}{\al_c}\frac{d\al_c}{d\be'}                         \com  \nn\\
\mbox{Free Energy}       &
-\Ga^{eff}|_c=\ -S_{\la}[\vp_c]-\la_c A\com \nn
\end{eqnarray}
%%%%%%%%%%%%%%%%%%%%%%%%%%%%%
where $w\equiv 16\pi\be'\ga$\ and $C(A)$\ is given in (\ref{3.11})\cite{N3b}
%\footnote
%{Because of the term $-\la (\int d^2x~e^{\vp}-A)$\ in $Y[A,\la]$,(\ref{3.4f}),
%we see $\la$~can be interpreted as the string (surface) tension.
%}
{}.
Note that $\Xi=0$.

$\qq$ The curvature$\times A$\
(\ $=R(\vp_c)\times A=\al(w)$\ ),
the classical value of $A<\intx \sqg R^2>$\ (\ =
$-\frac{\pl \Ga^{eff}[\vp_c]}{\pl\be'}$)
, the string tension$\times \ga A\ =\ga\la_c A\ $ and
the total free energy $\times \ga$\ ($=-\ga\Ga^{eff}[A]
=-\ga (S_{\la}+\la_c A)$\ )
are plotted, for the $-$branch solution,
in Fig.3 ,4,5 and 6 respectively.
In the figures we take $\xi=0.99$~whose meaning will be explained in Sec.5,
and the curves for the negative curvature solution (see Sec.3.2) are
also plotted.
The asymptotic behaviours of these quantities will be listed in Sec.3.3.
We also plot the area
$\times \frac{1}{A}$\ (
$=\frac{1}{A}\intx e^{\vp_c}=\frac{8\pi}{\al}$\ ) as the function of $w$
in Fig.7.
It shows
,as far as the classical configuration is concerned,
the $\del$-function condition in (\ref{3.3})
 is not satisfied except
for the $\al^+_c$-solution in the $\be$~(or $w$~) $\rightarrow +\infty$ region
where $\la\rightarrow +\infty$\ ,
and for the $\al^-_c$-solution in the $\be$~(or $w$~)
$\rightarrow -\infty$ region
where $\la\rightarrow -\infty$\ .
This has happened because we are approximating
the quantumly fluctuating manifold
by the simple classical sphere whose configuration is specified
only by the effective area
$\frac{1}{\al}$\ and the string tension $\la$\ .
This characteristically shows the present effective action approach using
$Y[A,\la]$\ (\ref{3.4f}).

{\vs 5}
\begin{center}
Fig.3\q $A\times$ Curvature $=\al(w)$\ ,\
Positive and Negative Curv. Sols., $-$Branch
\end{center}
{\vs 5}
\begin{center}
Fig.4a\q Log-Log Plot of $A<\intx\sqg R^2>|_c$, $w>0$,
Positive and Negative Curv. Sols., $-$Branch
\end{center}
{\vs 5}
\begin{center}
Fig.4b\q Linear Plot of $A<\intx\sqg R^2>|_c$,
Positive and Negative Curv. Sols., $-$Branch
\end{center}
{\vs 5}
\begin{center}
Fig.5\q$\ga A\times$(String Tension)$=\ga\la(w)A$,
Positive and Negative Curv. Sols., $-$Branch
\end{center}
{\vs 5}
\begin{center}
Fig.6\q$\ga \times$(Total Free Energy)$=-\ga\Ga^{eff}$,
Positive and Negative Curv. Sols., $-$Branch,
The $w$-independent terms ($C(A),\Ctil_2(A)$)
are omitted.
\end{center}
{\vs 5}
\begin{center}
Fig.7 \q $\frac{1}{A}\times$Area$=\frac{8\pi}{\al(w)}$,
Positive and Negative Curv. Sols., $-$Branch
\end{center}

%%%%%%%%%%%%%%%%%%%%%%%%%%%%%%%%%%%%%%%%%%%%%%%%%%%%%%%%%%%%%%%%%%%%%%
%%%%%%%%%%%%%%%%%%%%%%%%   3.2        %%%%%%%%%%%%%%%%%%%%%%%%%%%%%%%%
\subsection{Negative Curvature Solution}

$\qq$ For the case:
%*** 4.1    %%%%%%%%%%%%%%%%
\begin{eqnarray}
\al <0\com                                            \label{4.1}
\end{eqnarray}
%%%%%%%%%%%%%%%%%%%%%%%%%%%%%
the solution of (\ref{3.8}) is given by (cf.\cite{SEI}),
%*** 4.2    %%%%%%%%%%%%%%%%
\begin{eqnarray}
\vp_n(r;\al )=-ln~\{ \frac{-\al}{8}(1-\frac{r^2}{A})^2\}\pr  \label{4.2}
\end{eqnarray}
%%%%%%%%%%%%%%%%%%%%%%%%%%%%%
The equations (\ref{3.9a}-\ref{3.7}) and
(\ref{4.1}-\ref{4.2}) constitute another solution of (\ref{3.6a}).
It is singular at $r=\sqrt{A}$~, which means
the manifold is open. There exist two independent regions:\ the inner region
\ $0\leq r<\sqrt{A}$\ and the outer region\ $r>\sqrt{A}$. We consider only the
inner region\cite{N3c}
%\footnote
%{
%Adding the effect of the outer region does not change the essential part
%of the following content.
%}
{}.
We should carefully treat the singularity
by introducing a proper regularization. We regularize the inner region
by $0\leq r<\sqrt{A-\ep}$~ where $\ep\rightarrow +0$~. See Fig.8.

{\vs 5}
\begin{center}
Fig.8\ Reguralization of singularity at $r=\sqrt{A}$ \ of (\ref{4.2}).
\end{center}

Then various terms in the action are evaluated as
%*** 4.3    %%%%%%%%%%%%%%%%
\begin{eqnarray}
\intx~e^{\vp_n}
=\lim_{\ep\rightarrow +0}\int_{0\leq r^2 <A-\ep}
{}~d^2x~e^{\vp_n}
=\frac{8\pi}{-\al}A\lim_{\ep\rightarrow +0}
(\frac{A}{\ep}-1)\com                   \nn\\
\intx~\vp_n\pl^2\vp_n
=8\pi \lim_{\ep\rightarrow +0}\{2-(\frac{A}{\ep}-1)ln~(\frac{-\al}{8})
+\frac{2A}{\ep}(ln\frac{A}{\ep}-1)\ \}\com                  \label{4.3}\\
\intx~e^{-\vp_n}(\pl^2\vp_n)^2
=\frac{-8\pi\al}{A}\lim_{\ep\rightarrow +0}
(\frac{A}{\ep}-1)\com                 \nn\\
-\left.\intx\sqg R\right|_{\vp_n}=\intx~\pl^2\vp_n
=8\pi\lim_{\ep\rightarrow +0}(\frac{A}{\ep}-1)\com       \nn\\
\intx\pl_a\vp_n\pl_a\vp_n=16\pi
\lim_{\ep\rightarrow +0}(~ln\frac{\ep}{A}+\frac{A}{\ep}-1)\com\nn
\end{eqnarray}
%%%%%%%%%%%%%%%%%%%%%%%%%%%%%
where $\ep$ \ is introduced as a regularization parameter and is
a positive infinitesimally-small constant with the dimension of area
\cite{N3cc}.
The divergence of the total area $\intx~e^{\vp_n}$
, at this stage, says the manifold
considered is not closed. The singular point $r=\sqrt{A}$\ corresponds
to the boundary of the open manifold.

$\qq$ Using the above results, the Euclidean action (\ref{3.4g}) and
eq. (\ref{3.4l}), at the classical level, are given by
%*** 4.4    %%%%%%%%%%%%%%%%
\begin{eqnarray}
S_{\la}[\vp_n]=-\frac{4\pi}{\ga}(1+\xi)\La~ ln~(\frac{-\al}{8})
+8\pi\al\be'\La
+\frac{8\pi}{\al}\la A \La +C_2(A) \com                          \nn\\
C_2(A)=\frac{4\pi(1+\xi)}{\ga}\{ 2+2(\La+1)(ln(\La+1)-1)\}
+\frac{8\pi\xi}{\ga}(-ln(\La+1)+\La)\com\nn\\
\La\equiv \lim_{\ep\rightarrow +0}(\frac{A}{\ep}-1)>0\com   \nn\\
\frac{dS_{\la} [\vp_n]}{d\la}+A=
\{ -\frac{4\pi\La}{\ga\al}(1+\xi)+8\pi\be'\La-\frac{8\pi}{\al^2}\la A\La\}
\frac{d\al}{d\la}+\frac{8\pi}{\al}A\La+A                \label{4.4}\\
=\{-\frac{4\pi\La}{\ga\al}(1+\xi)+8\pi\be'\La-\frac{8\pi}{\al^2}\la A\La
+(2\al\be'-\frac{1}{\ga})(\frac{8\pi\La}{\al}+1)\}\frac{d\al}{d\la}=0
                                                        \pr\nn
\end{eqnarray}
%%%%%%%%%%%%%%%%%%%%%%%%%%%%%
The above equations are divergent and are not well-defined
for $\ep\rightarrow +0$.
%% %% 94.4.13 2:00AM
We can,however,absorb the divergence by the rescaling of the
coupling $\be'$\ ,the curvature parameter $\al$\ ,the cosmological
parameter $\la$\ and some physical operators to be evaluated.
%*** 4.5    %%%%%%%%%%%%%%%%
\begin{eqnarray}
\altil\equiv \frac{\al}{\La}\com\q \betil'\equiv\La\be'\ (\betil\equiv\La\be)
\com  \label{4.5}
\end{eqnarray}
%%%%%%%%%%%%%%%%%%%%%%%%%%%%%
where $\La$~is the positive divergent constant introduced in (\ref{4.4}).
Note that $\al$ \ does not change its sign by this transformation:\
$\altil<0$\ .The corresponding transformation of $\la$\ is obtained by
the requirement of keeping the form of (\ref{3.9a}).
%*** 4.6    %%%%%%%%%%%%%%%%
\begin{eqnarray}
\mbox{COND.}{\tilde 1}\qqqq
\latil A\equiv \frac{\la}{\La}A=\altil^2\betil'-\frac{\altil}{\ga}\pr
                                                             \label{4.6}
\end{eqnarray}
%%%%%%%%%%%%%%%%%%%%%%%%%%%%%

$\qq$ In terms of rescaled quantities, we can rewrite (\ref{4.4}) as
%*** 4.7    %%%%%%%%%%%%%%%%
\begin{eqnarray}
\Stil_{\latil}[\vp_n]\equiv\frac{S_{\la}[\vp_n]}{\La}
=-\frac{4\pi}{\ga}(1+\xi)~ln~(\frac{-\altil}{8})+16\pi\altil\betil'
+\Ctil_2(A)  \com \nn\\
\Ctil_2(A)=\frac{8\pi(1+\xi)}{\ga}ln~\La-\frac{8\pi}{\ga}  \com \label{4.7}  \\
\frac{dS_{\la} [\vp_n]}{d\la}+A=
\frac{d\Stil_{\latil}[\vp_n]}{d\latil}+A=
\frac{d\altil}{d\latil}[-\frac{4\pi}{\ga}\frac{1}{\altil}(1+\xi)+16\pi\betil'
+2\altil\betil'-\frac{1}{\ga}]=0\ .\nn
\end{eqnarray}
%%%%%%%%%%%%%%%%%%%%%%%%%%%%%
This result shows the rescaled action is finite except a constant term
(log-divergent) and the equation for $\latil_n$\ or $\altil_n$
\ (i.e.\ eq.(\ref{3.4l}))
, at the classical level, is completely
free from divergence.
%*** 4.8    %%%%%%%%%%%%%%%%
\begin{eqnarray}
\mbox{COND.}{\tilde 2} &{\tilde \Xi}(\altil;\betil',\ga,\xi)\equiv
2\betil'\altil^2+(16\pi\betil'-\frac{1}{\ga})\altil
-\frac{4\pi}{\ga}(1+\xi)=0\com  \nn\\
& \altil^{\pm}_n=\frac{1}{4\betil'}\{ -16\pi\betil'+\frac{1}{\ga}
\pm\sqrt{D   } ~\}
\com         \label{4.8}\\
  &
D\equiv (16\pi\betil')^2+\frac{1}{\ga^2}+32\pi\frac{\betil'}{\ga}\xi
=(16\pi\betil'+\frac{\xi}{\ga})^2+\frac{1}{\ga^2}(1-\xi^2)\pr   \nn
\end{eqnarray}
%%%%%%%%%%%%%%%%%%%%%%%%%%%%%
COND.${\tilde 1}$\ (\ref{4.6})\ gives
$\latil^\pm_n(\betil)=\latil(\betil,\altil^\pm_n(\betil))$.
We consider,again, the following region of $\xi$\ ,in order to
guarantee $D\geq 0$~ for all real $\betil$.
%*** 4.8b    %%%%%%%%%%%%%%%%
\begin{eqnarray}
-1\leq\xi\leq 1\q\pr        \label{4.8b}
\end{eqnarray}
%%%%%%%%%%%%%%%%%%%%%%%%%%%%%
The physical quantity of $<\intx~\sqg R^2>|_n$\ is given by
%*** 4.9    %%%%%%%%%%%%%%%%
\begin{eqnarray}
-\frac{A}{\La^2}<\intx\sqg R^2>|_{n}=
\frac{1}{\La^2}\frac{d\Ga^{eff}[\vp_n]}{d\be'}|_n
=\frac{d{\tilde \Ga}^{eff}[\vp_n]}{d\betil'}|_n                 \nn\\
=+16\pi\altil_n+\altil_n^2+
{\tilde \Xi}\times\frac{1}{\altil_n}\frac{d\altil_n}{d\betil}\com\q
                                                    \label{4.9}
\end{eqnarray}
%%%%%%%%%%%%%%%%%%%%%%%%%%%%%
where ${\tilde \Xi}=0$.

$\qq$ We note the difference in signs between the equations in the positive-
curvature case, (\ref{3.17}-\ref{3.18}), and those in the negative-curvature
case, (\ref{4.6}-\ref{4.9}).
Remarkably,by the following sign change,
%*** 4.10    %%%%%%%%%%%%%%%%
\begin{eqnarray}
\betil\rightarrow -\betil\com
\altil\rightarrow -\altil\com
(\q\latil\rightarrow -\latil\com
\Stil_{\latil}-\Ctil_2(A)\rightarrow -(\Stil_{\latil}-\Ctil_2(A))
\com\q)    \label{4.10}
\end{eqnarray}
%%%%%%%%%%%%%%%%%%%%%%%%%%%%%
the above negative-curvature results
($\altil_n,\Stil_{\latil}-\Ctil_2(A),\latil_n$)
as the functions of $\betil'$\ ,
reduce to  the same forms of the positive-curvature ones
($\al_c,S_{\la}-C(A),\la_c$) as the functions of $\be'$\cite{N3d}
%\footnote{
%The infrared regularization parts, $C(A)$~ and $\Ctil_2(A)$~, does not
%obey the reflection rule. Those parts do not depend on $w$~ and
%directly come from
%the topology of the manifold. The toplogy of a sphere and that of
%the infrared-regularized sphere (Fig.9) is quite different.
%}
{}.

$\qq$ We show the behaviours of $\altil^{\pm}_n,
-\frac{\pl{\tilde \Ga}^{eff}_{\pm}}{\pl\betil'}$\ , $\ga\latil^{\pm}_n A$\
and $-\ga{\tilde \Ga}^{eff}_\pm (\ =-\ga(\Stil^\pm_\latil +\latil^\pm_n A)\ )$
\
in the dotted lines of
Fig.3,4,5 and 6 respectively. The figures show the above reflection
symmetry clearly.
The asymptotic behaviours of the above physical quantities will be
listed in Table 3 of Sec.3.3.

It is very interesting that we can define finite quantities
in the open manifold in the above rescaling procedure
(\ref{4.5}-\ref{4.7}). It reminds us of the renormalization in the
quantum field theory. The present case is, however, a procedure to absorb
the infrared divergence due to the coordinate singularity of the classical
open manifold, not to absorb the ultraviolet divergence in the quantum theory.
Note that the constant-curvature sign remains negative after the rescaling
and the Euler number $\intx\sqg R$~ is negatively divergent.
These facts make us envisage Fig.9 as the rescaled manifold.
It describes a sphere punctured over the surface. Each puncture absorbs
the infrared divergence.

{\vs 5}
\begin{center}
Fig.9\ Punctured sphere absorbing infrared divergence
\end{center}

%%%%%%%%%%%%%%%%%%%%%%%%%%%%%%%%%%%%%%%%%%%%%%%%%%%%%%%%%%%%%%%%%%%%%%
\subsection{ Phases and Asymptotic Behaviours}

$\qq$ The asymptotic behaviours of the physical quantites,obtained in Sec.3.1,
are listed in Table 2, where the case of $\al<0$~
is excluded due to the present condition (\ref{3.10b}).

\vspace{0.5cm}
%%%%%%%%%%%%%%%%%%%%%%%%%%%%%%%%%%%%%%%%%%%%%%%%%%%%%%%%%%%%%%%%%%%%%%%%%%
%%%%%%%%%%%%%%%%%%%%%  Table 2   %%%%%%%%%%%%%%%%%%%%%%%%%%%%%%%%%%%%%%%%%
%%%%%%%%%%%%%%%%%%%%%%%%%%%%%%%%%%%%%%%%%%%%%%%%%%%%%%%%%%%%%%%%%%%%%%%%%%
\begin{tabular}{|c|c|c|c|c|c|}
\hline
  &  & $w\ll -1$  & $-1\ll w<0$
         & $0<w\ll 1$
            & $1\ll w$          \\
\hline
 & Phase & /&/&(E)&(D)                 \\
\cline{2-6}
  & $\al^+_c$ &$<0$\ ,
     & $<0$\ ,
         & $\frac{4\pi}{w}\{2+w(1-\xi)$
            & $4\pi\{2+\frac{1-\xi}{w} $    \\
  & &\ not allowed & \ not allowed
         &  $+O(w^2)\}$ & $+O(w^{-2})\}$    \\
\cline{2-6}
$+$  & $-\frac{\pl \Ga^{eff}_+}{\pl\be'}$ &/
     & / & $-\frac{64\pi^2}{w^2}\{ 1-(1+\xi)w$
                       & $64\pi^2\{1+\frac{0}{w}$          \\
  &  &  & &$\ \ +O(w^2)\}$  & $+O(w^{-2})\}$              \\
\cline{2-6}
  & $\ga\la^+_cA$&/
      & / & $\frac{4\pi}{w}\{-1+0\cdot w$
             & $4\pi w\{1$     \\
& & & & $+O(w^2)\}$ & $+O(w^{-1})\}$       \\
\cline{2-6}
 & & /&/& $\frac{4\pi}{w}\{1+2w$ & $4\pi w\{1 $      \\
 & $-\ga \Ga^{eff}_+$ &  & & $+(1+\xi)w~ln~w$ & $+O(w^{-1})\} $  \\
 & & & & $+O(w^2)\}-\ga C(A)$ & $-\ga C(A) $          \\
\hline
 & Phase & (C) & \multicolumn{2}{c|}{(B)} & (A)                             \\
\cline{2-6}
  & $\al^-_c$ & $8\pi$
     & \multicolumn{2}{c|}{ $4\pi(1+\xi)\{1$ }
            & $\frac{4\pi(1+\xi)}{w}$                \\
& & $+O(|w|^{-1})\}$ & \multicolumn{2}{c|}{$-\frac{1-\xi}{2}w\}+O(w^2)$}
            & $+O(w^{-2})$                                  \\
\cline{2-6}
$-$  & $-\frac{\pl \Ga^{eff}_-}{\pl\be'}$ & $64\pi^2+\frac{0}{|w|}$
     & \multicolumn{2}{c|}{
                   $16\pi^2(1+\xi)\{3-\xi$  }
            & $\frac{64\pi^2(1+\xi)}{w} $               \\
 & & $+O(w^{-2})$
       &  \multicolumn{2}{c|}{
            $-(1-\xi)^2w\}+O(w^2)$ }
                  & $+O(w^{-2})$                          \\
\cline{2-6}
 & $\ga\la^-_cA$ &$-4\pi |w|\{1$
     & \multicolumn{2}{c|}{ $4\pi(1+\xi)\{-1$  }
            & $-\frac{\pi}{w}(1+\xi)(3-\xi)$          \\
 & & $+O(\frac{1}{|w|})\}$
     & \multicolumn{2}{c|}{$+\frac{3-\xi}{4}w\}+O(w^2)$}
             & $+O(w^{-2}) $                               \\
\cline{2-6}
 & & $-4\pi |w|\{1$ & \multicolumn{2}{c|}{
              $4\pi(1+\xi)\{1-ln~\frac{1+\xi}{2} $
                                          }
                             & $4\pi(1+\xi)~ln~w$   \\
 & $-\ga\Ga^{eff}_-$ & $+O(\frac{1}{|x|})\} $
               & \multicolumn{2}{c|}{ $+\frac{3-\xi}{4}w\}+O(w^2)$ }
                        & +const                \\
 & & $-\ga C(A)$ & \multicolumn{2}{c|}{$-\ga C(A)$}  & $-\ga C(A) $ \\
\hline
\multicolumn{6}{c}{\q}                                   \\
\multicolumn{6}{c}{Table 2\ \  Asymp. behaviour of physical quantities.}\\
\multicolumn{6}{c}{
$R>0, w\equiv 16\pi\be'\ga, \ga=\frac{48\pi}{26-c_m}>0\ (c_m<26)$.
$C(A)$ is given by (\ref{3.11}).}
\end{tabular}
%%%%%%%%%%%%%%%%%%%%%%%%%  END  of  Table 1 %%%%%%%%%%%%%%%%%%%%%%%%%%%%%
%%%%%%%%%%%%%%%%%%%%%%%%%%%%%%%%%%%%%%%%%%%%%%%%%%%%%%%%%%%%%%%%%%%%%%%%%
\vspace{0.5cm}

$\qq$ Due to the 'reflection symmetry',
each phase of the negative-curvature solution,
given in Sec.3.2, is characterzed
in the similar way as in the positive-curvature case.
We list the phase characterization in Table 3.
('Primes' in Table 3 mean modification due to the sign difference.)

%%%%%%%%%%%%%%%%%%%%%%%%%%%%%%%%%%%%%%%%%%%%%%%%%%%%%%%%%%%%%%%%%%%%%%%%%
%%%%%%%%%%%%%%%%%%%%%  Table 3   %%%%%%%%%%%%%%%%%%%%%%%%%%%%%%%%%%%%%%%%%
%%%%%%%%%%%%%%%%%%%%%%%%%%%%%%%%%%%%%%%%%%%%%%%%%%%%%%%%%%%%%%%%%%%%%%%%%%
\begin{tabular}{|c|c|c|c|c|}
\hline
  & $w\ll -1$  & $-1\ll w<0$
         & $0<w\ll 1$
            & $1\ll w$          \\
\hline
+ & (D') & (E') & / & /                 \\
\hline
- & (A') & \multicolumn{2}{c|}{(B')} & (C')                             \\
\hline
\multicolumn{5}{c}{\q}                                   \\
\multicolumn{5}{c}{Table 3\q Phases of Negative Curvature Solution.
$w\equiv 16\pi\betil'\ga$.}
\end{tabular}
%%%%%%%%%%%%%%%%%%%%%%%%%  END  of  Table 3 %%%%%%%%%%%%%%%%%%%%%%%%%%%%%
%%%%%%%%%%%%%%%%%%%%%%%%%%%%%%%%%%%%%%%%%%%%%%%%%%%%%%%%%%%%%%%%%%%%%%%%%

\vspace{0.5cm}
$\qq$ All phases are explained in \cite{ITY1}
using the above asymptotic behaviour.
In the present paper we will characterize each phase by the field equation
satisfied in each asymptotic region in Sec.4 and by
the equation of state in Sec.7.
\vspace{1cm}
%%%%%%%%%%%%%%%%%%%%%%%%%%%%%%%%%%%%%%%%%%%%%%%%%%%%%%%%%%%%%%%%%%%%%%%%%
%%%%%%%%%%%%%%%%%%%%%%%%%%%%%%%%%%%%%%%%%%%%%%%%%%%%%%%%%%%%%%%%%%%
%%%%%%%%%%%%%%%%%%%%%%%%%%  4  %%%%%%%%%%%%%%%%%%%%%%%%%%%%%%%%%%%%
%%%%%%%%%%%%%%%%%%%%%%%%%%%%%%%%%%%%%%%%%%%%%%%%%%%%%%%%%%%%%%%%%%%
%%%%%%%%%%%%%%%%%%%%%%%%%%%%%%%%%%%%%%%%%%%%%%%%%%%%%%%%%%%%%%%%%%%
%%%%%%%%%%%%%%%%%%%%%%%%%%%%%%%%%%%%%%%%%%%%%%%%%%%%%%%%%%%%%%%%%%%%%
\section{Asymptotic Regions}
$\qqq$Now we consider the classical solutions of (\ref{3.6a}) in the
asymptotic regions\ :\
(a) $|w|\rightarrow \infty$\ ;\
(b) $|w|\rightarrow +0$\ .
Table 2 and 3 in Sec.3.3 say each region has  two cases.

%%%%%%%%%%%%%%%%%%%%%%%%%%  (ai) %%%%%%%%%%%%%%%%%%%%%%%%%%%%%%%%%%%
\flushleft{(ai)\ $|\ga\la A|\sim O(\frac{1}{w})$\ ,\
Weak-Field Vacua, ((A),(A'))}

In this region, the following parts of (\ref{3.6a}) are dominant.
%**** 4.3.5    %%%%%%%%%%%%%%%%
\begin{eqnarray}
 e^{-\vp}(\pl^2\vp)^2-2\pl^2(e^{-\vp}\pl^2\vp)=0\pr\nn\\
\vp_{ary}=\mbox{const}\com\ R(\vp_{asy})=0\pr
                                                         \label{4.3.5}
\end{eqnarray}
%%%%%%%%%%%%%%%%%%%%%%%%%%%%%
These vacua are defined only by the 'kinetic terms' in the action.
Therefore we call these vacua ((A),(A'))
{\it Weak-Field}(WF-){\it Vacua}\cite{N4a}.
%\footnote{
%Note that these vacua are meaningful only locally.
%The global constraint $\intx\sqg R=8\pi$\ is violated in these
%vacua and
%is irrelevent for the dynamics.
%}
%%%%%%%%%%%%%%%%%%%%%%%%%%%%%%%%% (aii) %%%%%%%%%%%%%%%%%%%%%%%%%%%%%%%
\flushleft{(aii)\ $\ga\la A\sim O(w)$\ ,\ Perfect Sphere Vacua,
((C),(C'),(D),(D'))}
%**** 4.3.6    %%%%%%%%%%%%%%%%
\begin{eqnarray}
 e^{-\vp}(\pl^2\vp)^2-2\pl^2(e^{-\vp}\pl^2\vp)-c~e^\vp=0\pr\nn\\
-R|_{asy}=e^{-\vp_{asy}}\pl^2\vp_{asy}=\mbox{const}=\pm\sqrt{c}(\neq 0)\pr
                                                         \label{4.3.6}
\end{eqnarray}
%%%%%%%%%%%%%%%%%%%%%%%%%%%%%
For the positive curvature case (lower sign case), $c$\ can be fixed
,by the condition $\intx\sqg R=8\pi$\ ,as $\sqrt{c}=\frac{8\pi}{A}$\
and the total area is $\frac{8\pi}{R}=A$\ .
These vacua
are strongly restricted by the 'potential term' of $e^\vp$\ and
describe a perfect sphere.
We call the vacua (D) and (C') {\it expansive perfect sphere}
where the string tension positively divergent, and
the vacua (C) and (D') {\it tensed perfect sphere}
where the string tension negatively divergent.

%%%%%%%%%%%%%%%%%%%%%%%%%%%%% (bi) %%%%%%%%%%%%%%%%%%%%%%%%%%%%%%%
\flushleft{(bi)\ $\ga\la A\sim \mbox{const}$\ ,
 Liouville Vacua, ((B),(B'))}
%**** 5b.1    %%%%%%%%%%%%%%%%
\begin{eqnarray}
\frac{1}{\ga}\pl^2\vp-\la(0)~e^\vp =0\com\q\mbox{Liouville Eq.}\q\com\nn\\
R=-e^{-\vp}\pl^2\vp=-\ga\la(0)\pr
                                                   \label{5b.1}
\end{eqnarray}
%%%%%%%%%%%%%%%%%%%%%%%%%%%%%
This  corresponds to the $\be=0$~theory.
In Phase (B), the Euler number is properly given by
$\intx\sqg R=-\ga\la(0)\cdot\intx\sqg=-\ga\la(0)A\cdot\frac{8\pi}{\al(0)}
=4\pi(1+\xi)\cdot\frac{8\pi}{4\pi(1+\xi)}=8\pi$~
(for arbitrary $\xi$~).
We call these regions Liouville Vacua.

%%%%%%%%%%%%%%%%%%%%%%%%%%%%%%% (bii) %%%%%%%%%%%%%%%%%%%%%%%%%%%
\flushleft{(bii)\ $\ga\la A\sim O(\frac{1}{w})$\ ,
Degenerate Vacua, ((E),(E'))}

In these regions the curvature must depend on $w$~ in order that
Eq.(\ref{3.6a}) is satisfied.
%**** 5b.2    %%%%%%%%%%%%%%%%
\begin{eqnarray}
\frac{1}{\ga}e^{-\vp}\pl^2\vp
+\be\{ (e^{-\vp}\pl^2\vp)^2-2e^{-\vp}\pl^2(e^{-\vp}\pl^2\vp)\}-\la(\be) =0\com
                                                             \nn\\
R=-e^{-\vp}\pl^2\vp\sim\ \frac{1}{A}\times O(\frac{1}{w})\pr\label{5b.2}
\end{eqnarray}
%%%%%%%%%%%%%%%%%%%%%%%%%%%%%
All terms of (\ref{3.6a}) are effective.
Because the total area $\frac{8\pi}{R}$~ vanishes, we name these regions
{\it degenerate vacua}.

\vspace{0.5cm}
We list all above asymptotic regions in Table 3 with the effective terms
of (\ref{3.6a}) marked by $\bigcirc$.

%%%%%%%%%%%%%%%%%%%%%%%%%%%%%%%%%%%%%%%%%%%%%%%%%%%%%%%%%%%%%%%%%%%%%%%%%%
%%%%%%%%%%%%%%%%%%%%%  Table 3   %%%%%%%%%%%%%%%%%%%%%%%%%%%%%%%%%%%%%%%%%
%%%%%%%%%%%%%%%%%%%%%%%%%%%%%%%%%%%%%%%%%%%%%%%%%%%%%%%%%%%%%%%%%%%%%%%%%%
\begin{tabular}{|c|c|c|c|c|}
\hline
\ \ Terms of (\ref{3.6a}) & $\frac{1}{\ga}\pl^2\vp$ &
$+\be\{ e^{-\vp}(\pl^2\vp)^2-2\pl^2(e^{-\vp}\pl^2\vp)\}$ &
                                  $-\la e^\vp $ &       \\
\hline
Weak-Field & - & $\bigcirc$ & - & (ai)       \\
\hline
Perf.Sphere & - & $\bigcirc$ & $\bigcirc$ & (aii) \\
\hline
Liouville  & $\bigcirc$ & - & $\bigcirc$ & (bi)  \\
\hline
Degenerate & $\bigcirc$ & $\bigcirc$ & $\bigcirc$ & (bii) \\
\hline
Free Boson & $\bigcirc$  & - & - & non-exist \\
\hline
           & $\bigcirc$  & $\bigcirc$ & - & non-exist \\
\hline
\multicolumn{5}{c}{\q}                                   \\
\multicolumn{5}{c}{Table 3\q Asymptotic States }
\end{tabular}
%%%%%%%%%%%%%%%%%%%%%%%%%  END  of  Table 3 %%%%%%%%%%%%%%%%%%%%%%%%%%%%%
%%%%%%%%%%%%%%%%%%%%%%%%%%%%%%%%%%%%%%%%%%%%%%%%%%%%%%%%%%%%%%%%%%%%%%%%%

\vspace{1cm}
Now we have characterized all asymptotic regions. We can see,
as shown in Fig.3,
$\al^-_c$ solution connects between WF-vacuum at $w=+\infty$\
and the tensed perfect sphere vacuum at $w=-\infty$. And
$\al^-_n$ solution connects between WF-vacuum at $w=-\infty$\
and the expansive perfect sphere vacuum at $\be=+\infty$.

%%%%%%%%%%%%%%%%%%%%%%%%%%%%%%%%%%%%%%%%%%%%%%%%%%%%%%%%%%%%%%%%%%%%%%%%%
%%%%%%%%%%%%%%%%%%%%%%%%%%%%%%%%%%%%%%%%%%%%%%%%%%%%%%%%%%%%%%%%%%%
%%%%%%%%%%%%%%%%%%%%%%%%%%  5  %%%%%%%%%%%%%%%%%%%%%%%%%%%%%%%%%%%%
%%%%%%%%%%%%%%%%%%%%%%%%%%%%%%%%%%%%%%%%%%%%%%%%%%%%%%%%%%%%%%%%%%%
%%%%%%%%%%%%%%%%%%%%%%%%%%%%%%%%%%%%%%%%%%%%%%%%%%%%%%%%%%%%%%%%%%%
\section{ $\la$-integral }

In this section we do the $\la$-integral of (\ref{3.4k}) in the
lowest order. This part gives us some contribution to $Z[A]$.
We consider the positive curvature solution.
After splitting $\la$~ around $\la_c$~ :\ $\la=\la_c+\om$~
, the $\om$-integral part of
(\ref{3.4ll}) is approximated as
%***(la.1)%%%%%%%%%%%%%%%%%%%%
\begin{eqnarray}
& Z_{\om}[A]
\equiv\int d\om~exp~\{~\frac{1}{2}
\left.\frac{d^2\Gahat[\la]}{d\la^2}\right|_{\la_c}~\om^2
                        +O(\om^3) \ \}         & \nn\\
& \approx
\int d\om~exp~\{~\frac{1}{2}
\left.\frac{d^2S_\la}{d\la^2}\right|_{\la_c}~\om^2\ \}\pr   &\label{la.1}
\end{eqnarray}
%%%%%%%%%%%%%%%%%%%%%%%%%%%%%
{}From (\ref{3.11}) and (\ref{3.9a}), we can obtain
%*** la.2    %%%%%%%%%%%%%%%%
\begin{eqnarray}
& \frac{d^2S_\la[\vp_c]}{d\la^2}=
-\frac{4\pi}{\ga}\frac{1+\xi}{\al^2}(\frac{d\al}{d\la})^2
+(\frac{4\pi}{\ga}\frac{1+\xi}{\al}
-16\pi\be')\frac{d^2\al}{d\la^2}                  & \nn\\
& =A^2\frac{4\pi}{\al^2}\frac{1}{(2\al\be'-\frac{1}{\ga})^3}
\{~4(1-\xi)\al^2\be'^2+(1+\xi)(2\al\be'-\frac{1}{\ga})^2~\}
                                                    & \label{la.2}\\
& =A^2\frac{4\pi}{\al^2}\frac{\ga}{(\frac{\al w}{8\pi}-1)^3}
\{~(1-\xi)(\frac{\al w}{8\pi})^2+(1+\xi)(\frac{\al w}{8\pi}-1)^2~\}\com &\nn
\end{eqnarray}
%%%%%%%%%%%%%%%%%%%%%%%%%%%%%
where we have used some relations derived from (\ref{3.9a})~:~
$d\al/d\la=A/(2\al\be'-\frac{1}{\ga})\com\
 d^2\al/d\la^2=-2\be'A^2/(2\al\be'-\frac{1}{\ga})^3\pr$\
Putting $\al^\pm_c$-solution of (\ref{3.18}) into the above expression,
we can confirm
$d^2S_\la/d\la^2|_{\al^-_c}< 0\ ,\
d^2S_\la/d\la^2|_{\al^+_c}> 0 $~ for all $\be$~(or $w$) region.
(Note that $\{\ \}$-part of (\ref{la.2}) is positive definite.)
For the $\al^-_c$-solution, it is necessary to
change the integeral path from the original pure imaginary $\om$~ in Fig.1
to the real $\om$~ as shown in Fig.10. $Z_\om[A]$~ is evaluated as
%**** la.3    %%%%%%%%%%%%%%%%
\begin{eqnarray}
& Z_\om[A]=\frac{1}{\sqrt{-\frac{d^2S_\la}{d\la^2}|_{\al^-_c}} }
\q \mbox{for $-$branch solution}                   & \nn\\
& Z_\om[A]=\frac{1}{\sqrt{+\frac{d^2S_\la}{d\la^2}|_{\al^+_c}} }
\q \mbox{for $+$branch solution}   &  \pr   \label{la.3}
\end{eqnarray}
%%%%%%%%%%%%%%%%%%%%%%%%%%%%%

{\vs 7}
\begin{center}
Fig.10\ $\la$-integral path for $-$branch solution
\end{center}

Using the results of Table 2, the asymptotic behaviours are evaluated as
%*** la.4    %%%%%%%%%%%%%%%%
\begin{eqnarray}
\mbox{Phase (A)}\q w\gg 1\com &
\ln~Z_\om\sim -\ln~A-\ln~w\com       \nn\\
\mbox{Phase (B)}\q |w|\ll 1\com &
\ln~Z_\om\sim -\ln~A+\mbox{const}\com       \nn\\
\mbox{Phase (C)}\q w\ll -1\com &
\ln~Z_\om\sim -\ln~A+\half\ln~|w|\com       \label{la.4}\\
\mbox{Phase (D)}\q w\gg 1\com &
\ln~Z_\om\sim -\ln~A+\half\ln~w\com       \nn\\
\mbox{Phase (E)}\q 0<w\ll 1\com &
\ln~Z_\om\sim -\ln~A+\half\ln~w\pr       \nn
\end{eqnarray}
%%%%%%%%%%%%%%%%%%%%%%%%%%%%%
We notice the first term of each right-hand side contributes to the
string susceptibility (see Sec.6). The second term does not
have a factor of $4\pi/\ga=(26-c_m)/12$~ in comparison with the
$\Ga^{eff}$~ of Table 2. Because the factor means the number of
freedom in this thermodynamical system (see Sec.7) , their contribution
is negligible except for the case :\ $c_m\approx 26$.
%%%%%%%%%%%%%%%%%%%%%%%%%%%%%%%%%%%%%%%%%%%%%%%%%%%%%%%%%%
%%%%%%%%%%%%%%%%%%%%%%%%%%%%%%%%%%%%%%%%%%%%%%%%%%%%%%%%%%%%%%%%%%%%%%
\vspace{1cm}
%%%%%%%%%%%%%%%%%%%%%%%%%%%%%%%%%%%%%%%%%%%%%%%%%%%%%%%%%%%%%%%%%%%
%%%%%%%%%%%%%%%%%%%%%%%%%%  6  %%%%%%%%%%%%%%%%%%%%%%%%%%%%%%%%%%%%
%%%%%%%%%%%%%%%%%%%%%%%%%%%%%%%%%%%%%%%%%%%%%%%%%%%%%%%%%%%%%%%%%%%
%%%%%%%%%%%%%%%%%%%%%%%%%%%%%%%%%%%%%%%%%%%%%%%%%%%%%%%%%%%%%%%%%%%
\section{ Cross-Over Points and Determination of $\xi$}
%%%%%%%%%%%%%%%%%%%%%%%%%%%%%%%%%%%%%%%%%%%%%%%%%%%%%%%%%%%%%%%%%%%%%%

$\qq$ Let us see the $\xi$-dependence of the cross-over points.
Because the negative constant curvature solution is obtained
by the reflection symmetry from the positive one, we discuss
only the latter one.
As for the +branch solution, the string tension
 $\la$\ changes
its sign at $w_0$\ ,which is located somewhere between
(D)-phase and (E)-phase of Table 2(see Fig.5)\ .
We can obtain it as the zero of
$\la^+_c(w_0)=0$\ in (\ref{3.18}).
%*** 3.19    %%%%%%%%%%%%%%%%
\begin{eqnarray}
w_0(\xi)=\frac{4}{3-\xi}\pr  \label{3.19}
\end{eqnarray}
%%%%%%%%%%%%%%%%%%%%%%%%%%%%%
The - solution has two cross-over points.
The log-log plot of
$-\frac{\pl \Ga^{eff}_-[\vp_c]}{\pl\be'}$ (Fig.4a) shows, at some point
$w_c>0$\ between phase (A) and (B), the behaviour changes from the
linearly-descending line to
the constant-line as we decrease $w$.
The linear plot of
$-\frac{\pl \Ga^{eff}_-[\vp_c]}{\pl\be'}$ (Fig.4b) shows, at some point
$w_c'<0$\ between phase (B) and (C), the behaviour changes from the
linearly-descending line to
the constant-line as we decrease $w$ in the negative region.
Those straight lines can be obtained as
%*** 3.20    %%%%%%%%%%%%%%%%
\begin{eqnarray}
-\frac{\pl \Ga^{eff}_-[\vp_c]}{\pl\be'}
\rightarrow 64\pi^2\frac{1+\xi}{w}\q\mbox{as}\q w\rightarrow +\infty
                                                    \com    \nn\\
-\frac{\pl \Ga^{eff}_-[\vp_c]}{\pl\be'}
\rightarrow 16\pi^2(1+\xi)\{(3-\xi)-(1-\xi)^2w
+O(w^2)\}\q\mbox{as}\q w\rightarrow +0\com            \label{3.20} \\
-\frac{\pl \Ga^{eff}_-[\vp_c]}{\pl\be'}
\rightarrow 64\pi^2+\frac{0}{|w|}+O(w^{-2})\q\mbox{as}\q w\rightarrow -\infty
                                                    \pr    \nn
\end{eqnarray}
%%%%%%%%%%%%%%%%%%%%%%%%%%%%%
We can clearly define the changing points $w_c$\ and $w_c'$\
as the cross-point of two corresponding
asymptotic lines above, and obtain as
%*** 3.21    %%%%%%%%%%%%%%%%
\begin{eqnarray}
w_c(\xi)=\frac{4}{3-\xi}= w_0(\xi)\com\q
w_c'(\xi)=-\frac{1}{1+\xi}\pr                 \label{3.21}
\end{eqnarray}
%%%%%%%%%%%%%%%%%%%%%%%%%%%%%

All cross-over points depend on
the parameter of the total derivative
term ,$\xi$\ ,and which says the global term controls the essential
behaviour of the theory.

$\qq$ What value should we take for $\xi$~? It can be answered
, purely within the theory, from the quantum analysis\cite{S1}.
When we take $\xi=1$~, the renormalization-group beta functions
have zeros for $w\geq 1$~. Here, however,
we fix the parameter $\xi$~ by adjusting the asymptotic
(\ $A\rightarrow\infty $\ )
behaviour of $Z[A]$, for the case $\be=0$\ ,
with the KPZ (conformal) result\cite{KPZ}. The asymptotic behaviour of $Z[A]$
for $\be=0$\ is given as
%*** 4.3.1    %%%%%%%%%%%%%%%%
\begin{eqnarray}
\al^-_c-\mbox{solution}                     \nn\\
Z[A]|_{w=0}\sim
A^{-\frac{8\pi\xi}{\ga}}\cdot A^{-1}=A^{-\frac{26-c_m}{6}\xi-1}\com\nn\\
\mbox{as}\ A\rightarrow +\infty\com
\label{4.3.1}
\end{eqnarray}
%%%%%%%%%%%%%%%%%%%%%%%%%%%%%
where the factor $A^{-1}$~ comes from $Z_\om$~ in Sec.5.
The KPZ result\cite{KPZ} is
%*** 4.3.2    %%%%%%%%%%%%%%%%
\begin{eqnarray}
Z^{KPZ}[A]\sim A^{\ga_s-3}\com\
\ga_s=\frac{1}{12}\{c_m-25-\sqrt{(25-c_m)(1-c_m)}\}+2\pr   \label{4.3.2}
\end{eqnarray}
%%%%%%%%%%%%%%%%%%%%%%%%%%%%%
In order to adjust our result with the KPZ result in the classical
limit $c_m\rightarrow -\infty$\ :\
$Z^{KPZ}[A]\sim A^{+\frac{1}{6}c_m}$\ , we must take
%*** 4.3.3    %%%%%%%%%%%%%%%%
\begin{eqnarray}
\xi=1\pr                \label{4.3.3}
\end{eqnarray}
%%%%%%%%%%%%%%%%%%%%%%%%%%%%%

Taking $\xi=1$, the asymptotic behaviour of $Z[A]$\ for the
$\al^-_c$-solution is
%*** 4.3.4    %%%%%%%%%%%%%%%%
\begin{eqnarray}
Z[A]\sim A^{-\frac{26-c_m}{6}-1}\com\q A\rightarrow +\infty
\pr  \label{4.3.4}
\end{eqnarray}
%%%%%%%%%%%%%%%%%%%%%%%%%%%%%
Now we compare the KPZ result and the semiclassical result in the
normalized form.
%*** 4.3.4b    %%%%%%%%%%%%%%%%
\begin{eqnarray}
Z^{KPZ}_{norm}[A]\equiv \frac{Z^{KPZ}[A]}{Z^{KPZ}[A]|_{c_m=0}}
\sim A^{\ga_s(c_m)-\ga_s(c_m=0)}\com\nn\\
\ga_s(c_m)-\ga_s(c_m=0)=\frac{1}{12}\{c_m+5-\sqrt{(25-c_m)(1-c_m)}\}
                                            \com   \label{4.3.4b}\\
Z_{norm}[A]\equiv \frac{Z[A]}{Z[A]|_{c_m=0}}
\sim A^{+\frac{c_m}{6}}\com\q \pr\nn
\end{eqnarray}
%%%%%%%%%%%%%%%%%%%%%%%%%%%%%

In Fig.11, the present semiclassical result and the KPZ result
are plotted.

{\vs 5}
\begin{center}
Fig.11\ Semiclassical result versus KPZ formula for string susceptibility
\end{center}

In the following, we take $\xi=1$\cite{N6a}
%\footnote{
%In the  text, we understand $\xi\rightarrow +1-0$.
%Practically all physical quantities ,in the text,
%are numerically evaluated at $\xi=0.99$.
%}
{}.

\vspace{1cm}
%%%%%%%%%%%%%%%%%%%%%%%%%%%%%%%%%%%%%%%%%%%%%%%%%%%%%%%%%%%%%%%%%%%
%%%%%%%%%%%%%%%%%%%%%%%%%%  7  %%%%%%%%%%%%%%%%%%%%%%%%%%%%%%%%%%%%
%%%%%%%%%%%%%%%%%%%%%%%%%%%%%%%%%%%%%%%%%%%%%%%%%%%%%%%%%%%%%%%%%%%
%%%%%%%%%%%%%%%%%%%%%%%%%%%%%%%%%%%%%%%%%%%%%%%%%%%%%%%%%%%%%%%%%%%
\section{ Phases, Thermodynamic Properties
 and Equation of State}
%%%%%%%%%%%%%%%%%%%%%%%%%%%%%%%%%%%%%%%%%%%%%%%%%%%%%%%%%%%%%%%%%%%%%%
%%%%%%%%%%%%%%%%%%%%%%%%%%%%%%%%%%%%%%%%%%%%%%%%%%%%%%%%%%%%%%%%%%%%%%%%%

$\qq$ In this section we examine the thermodynamic properties of the system
using the obtained analytic expression.
We consider the positive curvature solution.
The partition function is given by
%***(state.1)%%%%%%%%%%%%%%%%%%%%
\begin{eqnarray}
Z[A]=\int d\la~exp~\{~\Gahat[\la]+\la A~\}
\approx exp~\{\Gahat[\la_c]+\la_c A\}\com\nn\\
\frac{d}{d\la}(\Gahat[\la]+\la A)|_{\la_c}
=\frac{d\Gahat[\la_c]}{d\la_c}+A=0\pr   \label{state.1}
\end{eqnarray}
%%%%%%%%%%%%%%%%%%%%%%%%%%%%%
Under the variation of the total area:\ $A\ra A+\Del A$~,
$\ln~Z[A]$~ changes by
$\Del (\ln~Z[A])=\la_c\cdot\Del A+\Del A\cdot\frac{d\la_c}{dA}\cdot
(\frac{d\Gahat}{d\la}+A)|_{\la_c}=\la_c\cdot\Del A$~.
Because the free energy $F$~ is given by
$F=- ln~Z[A]$~, the pressure $P$~ is obtained as
%*** state.2    %%%%%%%%%%%%%%%%
\begin{eqnarray}
& P=-\frac{\pl}{\pl A}F=\frac{\pl}{\pl A}\ln~Z[A]=\la_c
      \pr &   \label{state.2}
\end{eqnarray}
%%%%%%%%%%%%%%%%%%%%%%%%%%%%%
The pressure is the same as the string tension.

$\qq$ We define the temerature $T(w)$~ , imitating the Boyle-Charles' law,
as follows.
%*** state.3    %%%%%%%%%%%%%%%%
\begin{eqnarray}
& P\cdot A\equiv \frac{4\pi}{\ga}~T(w) \com & \nn\\
& T(w)=\frac{\ga\la_c A}{4\pi}=\frac{1}{64\pi^2}({\al_c}^2 w-16\pi\al_c)
                                       \com & \label{state.3}\\
& \al_c(w)=\left\{ \begin{array}{ll}
\frac{4\pi}{w}\{ w+1~+|w-1| ~\} & \mbox{for + branch solution}\\
\frac{4\pi}{w}\{ w+1~-|w-1| ~\} & \mbox{for $-$branch solution}
               \end{array}
       \right.                            & \nn
\end{eqnarray}
%%%%%%%%%%%%%%%%%%%%%%%%%%%%%
$N\equiv 4\pi/\ga=(26-c_m)/12$~ corresponds to the 'mol number'.
The temperature is the (dimensionless) string tension per a unit mol.
The final analytic form of the temperature is given by
%*** state.4    %%%%%%%%%%%%%%%%
\begin{eqnarray}
+\mbox{branch solution}\qqq
& T(w)=\left\{ \begin{array}{ll}
-\frac{1}{w} & \mbox{for}\ 0<w\leq 1\\
w-2          & \mbox{for}\ 1\leq w
               \end{array}
       \right.                            & \nn\\
-\mbox{branch solution}\qqq
& T(w)=\left\{ \begin{array}{ll}
w-2            & \mbox{for}\ w\leq 1\\
-\frac{1}{w}   & \mbox{for}\ 1\leq w
               \end{array}
       \right.                            & \label{state.4}
\end{eqnarray}
%%%%%%%%%%%%%%%%%%%%%%%%%%%%%
The behaviour of $T(w)$~ is plotted in Fig.12.

{\vs 5}
\begin{center}
Fig.12\ Temperature $T=T(w)$, Pos. Curv. Sol.
\end{center}

$\qq$ The asymptotic form of temperature in each phase is given by,
%*** state.5    %%%%%%%%%%%%%%%%
\begin{eqnarray}
\mbox{Phase (A)}\q w\gg 1\com &
P\cdot A=-\frac{4\pi}{\ga}\frac{1}{w}\com &
T_{(A)}= -\frac{1}{w}\com       \nn\\
\mbox{Phase (B)}\q |w|\ll 1\com &
P\cdot A=-\frac{8\pi}{\ga}(1+O(w))\com &
T_{(B)}\approx -2\com       \nn\\
\mbox{Phase (C)}\q w\ll -1\com &
P\cdot A=\frac{4\pi}{\ga}w(1+O(w^{-1}))\com &
T_{(C)}\approx w\com       \label{state.5}\\
\mbox{Phase (D)}\q w\gg 1\com &
P\cdot A=\frac{4\pi}{\ga}w(1+O(w^{-1}))\com &
T_{(D)}\approx w\com       \nn\\
\mbox{Phase (E)}\q 0<w\ll 1\com &
P\cdot A=-\frac{4\pi}{\ga}\frac{1}{w}\com &
T_{(E)}= -\frac{1}{w}\pr       \nn
\end{eqnarray}
%%%%%%%%%%%%%%%%%%%%%%%%%%%%%
The negativeness of temperature, in Phase (A),(B) and (C),
says the matter-gass particles
attract each other.
The small absolute value of $T_{(A)}$~ says the matter-gass particles
 move almost freely.
We can do the same analysis for the negative curvature solution.
The corresponding temperture is obtained by reflecting the graph
of Fig.12 following (\ref{4.10}).
It is interesting that
the matter-gass particles atracting each other on an open manifold
can be regarded as the 'repulsive'
particles on a regularized closed manifold.

$\qq$ The entropy is similarly obtained. Using the relation:\
$\Del w|_{A:fixed}=w\cdot \frac{\Del\be}{\be}\com\q
\Del T|_{A:fixed}=\frac{\pl T}{\pl w}\cdot w\frac{\Del\be}{\be}$\ ,
it is given as
%*** state.6    %%%%%%%%%%%%%%%%
\begin{eqnarray}
& S_{ent}=-(\frac{\pl F}{\pl T})_A  &\nn\\
&= +\frac{1}{w}\frac{\pl w}{\pl T}\cdot \be\frac{\pl}{\pl\be}\ln~Z[A]
=-\frac{1}{16\pi\ga}\frac{\pl w}{\pl T}\cdot A<\intx\sqg R^2>\pr &
                                           \label{state.6}
\end{eqnarray}
%%%%%%%%%%%%%%%%%%%%%%%%%%%%%
We see the entropy is related to the expectation value:\
$A<\intx\sqg R^2>$~ considered in Sec.3, as above.
Using the follwoing results from (\ref{3.18}) ($\xi=1$~is taken),
%*** state.7    %%%%%%%%%%%%%%%%
\begin{eqnarray}
+\mbox{branch solution}\qqq
& \expect=\left\{ \begin{array}{ll}
(8\pi)^2(\frac{2}{w}-\frac{1}{w^2}) & \mbox{for}\ 0<w\leq 1\\
+(8\pi)^2                           & \mbox{for}\ 1\leq w
               \end{array}
          \right. \com                            & \nn\\
-\mbox{branch solution}\qqq
& \expect=\left\{ \begin{array}{ll}
+(8\pi)^2                            & \mbox{for}\ w\leq 1\\
(8\pi)^2(\frac{2}{w}-\frac{1}{w^2})  & \mbox{for}\ 1\leq w
               \end{array}
       \right. \com                            & \label{state.7}
\end{eqnarray}
%%%%%%%%%%%%%%%%%%%%%%%%%%%%%
we obtain the expression for the entropy.
%*** state.8    %%%%%%%%%%%%%%%%
\begin{eqnarray}
+\mbox{branch solution}\qqq
& S_{ent}=\left\{ \begin{array}{ll}
-\frac{4\pi}{\ga}(2w-1) & \mbox{for}\ 0<w\leq 1\\
-\frac{4\pi}{\ga}       & \mbox{for}\ 1\leq w
               \end{array}
       \right. \com                            & \nn\\
-\mbox{branch solution}\qqq
& S_{ent}=\left\{ \begin{array}{ll}
-\frac{4\pi}{\ga}        & \mbox{for}\ w\leq 1\\
-\frac{4\pi}{\ga}(2w-1)  & \mbox{for}\ 1\leq w
               \end{array}
       \right.     \pr                       & \label{state.8}
\end{eqnarray}
%%%%%%%%%%%%%%%%%%%%%%%%%%%%%
The graph of $S_{ent}$~ is plotted in Fig.13.

{\vs 5}
\begin{center}
Fig.13\ Entropy per unit mol, $S_{ent}(w)/(4\pi/\ga)$, Pos.Curv.Sol.
\end{center}

The largeness of the absolute value of $S_{ent}$~ in Phase (A) shows
the much amount of freedom of the system, whereas the fixed value
in Phase (B) and (C) shows the possible configurations are
restricted.

$\qq$ From the behaviours of the temperature and the entropy, the cross-over
in the $-$~solution looks to occur
only at one point, $w=1$~, on the w-axis. The corresponding one to
$w_c'$~ in Sec.6 does not appear.

$\qqq$In Fig.14, all phases above are pictorially depicted.
\vspace{10cm}
\begin{center}
Fig.14\ Schematic image of surface in each phase
\end{center}

%%%%%%%%%%%%%%%%%%%%%%%%%%%%%%%%%%%%%%%%%%%%%%%%%%%%%%%%%%%%%%%%%%%%%
%%%%%%%%%%%%%%%%%%%%%%%%%%%%%%%%%%%%%%%%%%%%%%%%%%%%%%%%%%%%%%%%%%%
%%%%%%%%%%%%%%%%%%%%%%%%%%  8  %%%%%%%%%%%%%%%%%%%%%%%%%%%%%%%%%%%%
%%%%%%%%%%%%%%%%%%%%%%%%%%%%%%%%%%%%%%%%%%%%%%%%%%%%%%%%%%%%%%%%%%%
\section{Discussions and Conclusions}
$\qq$ Among two branches ,the $-$ branch (of the positive curvature) solution
appears in the lattice simulation\cite{ITY1,ITY2}.
It is consistent with the present analysis, where $-$ branch
is energetically prefarable to + branch.
Some features of + branch are the same as those obtained in \cite{KN}
using the conformal field approach\cite{ITY1}. It seems important
to analyse the relation between the present semiclassical approach
and the conformal field approach.

$\qq$ We discuss the meaning and the possible
role of the negative curvature solution.
The existance of the  vacua related by the reflection symmetry:\
$R\change -R$\ , is one of stressing points of this paper.
We may say
the appearance of 'dual' solutions reflects the reflection symmetry\ :\
$R\change -R$\  in the 'induced' $R^2$-gravity\
$\Lcal_{ind}=\frac{1}{2\ga}R\frac{1}{\Del}R-\be R^2-\mu$\ .
The symmetry appears manifestly due to the $R^2$-term.
Their topologies, however, are different
:\ the positive curvature solution satisfies
\ $\intx\sqg~R=8\pi$\ , which means the sphere topology, whereas
the negative one satisfies
:\ $\intx\sqg~R=-8\pi\frac{A}{\ep}=-\infty$\ ,\ which means
the toplogy of a sphere with the infinite number of punctures(Fig.9).
We suppose
this reflection symmetry of vacua
is general for manifolds with other topologies.
It means a physical quantity on a manifold can also be calculated on another
manifold with a different topology. It requires further analysis for clarity.

$\qq$ The semiclassical approach can easily provide the physical meaning
such as thermodynamic properties. The pesent system can be regarded as
the closed thermodynamic system where many scalar-matter particles move
in the gravitational potential and whose configuration is thermally
in an equilibrium state. The $R^2$~ coupling, $w$~(or $\be$~), parametrises
the temerature. The phase difference can be thermodynamically
interpreted as the difference of $w$-dependence of the temperature.

$\qq$ The important role of the integration parameter $\la$~ introduced
in (\ref{3.4f}) and of the total derivative term discussed in Sec.5 show
the proper treatment of the area constraint and the topology
constraint is so important to understand the 2d QG.  In the treatment,
the infrared regularizations of Fig.2 and of Fig.8 are nicely used.

Evaluation of
the quantum effects to the present classical results is an important
work to be done. It can be taken into account  perturbatively as explained
in (\ref{3.4j}). The renormalization has already been analysed in \cite{S1}.

$\qq$ The present approach can be valid for the higher-dimensional quantum
gravity. The 3 dim QG has been recently 'measured' in the Lattice
simulation with a high statistics.   %*cite{Hagura et al}.
The semiclassical analysis of the data
will soon become  an urgent work to be done.
The success of the perturbative 2d QG
using the semiclassical
method is strongly encouraging for the further progress of the perturbative
quantum gravity in the realistic dimensions.

%%%%%%%%%%%%%%%%%%%%%%%%%%%%%%%%

\begin{flushleft}
{\bf Acknowledgement}
\end{flushleft}
The author thanks
N. Ishibashi, H. Kawai, N. Tsuda and T. Yukawa  for discussions
and comments about the present work.
%%%%%%%%%%%%%%%%%%%%%%%%%%%%%%%%%%%%%%%%%%%%%%%%%%%%%%%%%%%%%%%%%%%
%%%%%%%%%%%%%%%%%%%%%%%%%%  App. A    %%%%%%%%%%%%%%%%%%%%%%%%%%%%%
%%%%%%%%%%%%%%%%%%%%%%%%%%%%%%%%%%%%%%%%%%%%%%%%%%%%%%%%%%%%%%%%%%%
{\vs 2}
\appendix
\begin{flushleft}
{\Large\bf Appendix A. Some useful formula}
\end{flushleft}

Some useful formulas are collected here.
\begin{flushleft}
{(i) Positive curvature solution}
\end{flushleft}
%*** a.1    %%%%%%%%%%%%%%%%
\begin{eqnarray}
\vp_c(r;\al )=-ln~\{ \frac{\al}{8}(1+\frac{r^2}{A})^2\}\com\ \
                                             \al>0\pr  \label{a.1}
\end{eqnarray}
%%%%%%%%%%%%%%%%%%%%%%%%%%%%%
%*** a.2    %%%%%%%%%%%%%%%%
\begin{eqnarray}
e^{-\vp_c}=\frac{\al}{8}(1+\frac{r^2}{A})^2\com
\pl^2\vp_c=-\frac{8A^{-1}}{(1+A^{-1}r^2)^2}\com              \nn\\
\left.\intx\sqg\right|_{\vp_c}=\intx~e^{\vp_c}=\frac{8\pi}{\al}A\com
\intx~\vp_c\pl^2\vp_c=8\pi (2+~ln~\frac{\al}{8}\ )\com        \nn\\
\int_{x^2\leq L^2} d^2x\pl_a\vp_c\pl_a\vp_c=
16\pi\{~ln(1+\frac{L^2}{A})-\frac{L^2/A}{1+L^2/A}~\}\com      \nn\\
\left.\intx\sqg R^2\right|_{\vp_c}=\intx~e^{-\vp_c}(\pl^2\vp_c)^2
=\frac{8\pi\al}{A}\com                                      \label{a.2}\\
-\left.\intx\sqg R\right|_{\vp_c}=\intx~\pl^2\vp_c=-8\pi\com  \nn\\
(\q\mbox{cf.\ \ For the oriented closed manifold:\ \ }
\intx\sqg R=8\pi(1-h)\ ,\ \                                  \nn\\
h=\mbox{Number of handles}\q)\com   \nn\\
\left.R\right|_{\vp_c}=-e^{-\vp_c}\pl^2\vp_c=\frac{\al}{A}>0\pr  \nn
\end{eqnarray}
%%%%%%%%%%%%%%%%%%%%%%%%%%%%%
%**** a.3   %%%%%%%%%%%%%%%%
\begin{eqnarray}
2A^{-1}\int_0^{\infty}\frac{r~dr}{(1+A^{-1}r^2)}=1\com
\int\frac{ln~(x+1)}{(x+1)^2}=-\frac{ln~(x+1)}{x+1}-\frac{1}{x+1} \pr
                                                             \label{a.3}
\end{eqnarray}
%%%%%%%%%%%%%%%%%%%%%%%%%%%%%
\begin{flushleft}
{(ii) Negative curvature solution}
\end{flushleft}
%*** a.4    %%%%%%%%%%%%%%%%
\begin{eqnarray}
\vp_n(r;\al )=-ln~\{ \frac{-\al}{8}(1-\frac{r^2}{A})^2\}\com\ \
                                             \al<0\pr  \label{a.4}
\end{eqnarray}
%%%%%%%%%%%%%%%%%%%%%%%%%%%%%
This solution is singular at $r=\sqrt{A}$. The following volume
integrals are evaluated using two ways of regularization:\
$k=1$) $\intx=\lim_{\ep\rightarrow +0}\int_{0\leq r^2\leq A-\ep}$~and
$k=2$) $\intx=\lim_{\ep\rightarrow +0}(\int_{0\leq r^2\leq A-\ep}
                                          +\int_{A+\ep\leq r^2})$~.

%*** a.5    %%%%%%%%%%%%%%%%
\begin{eqnarray}
e^{-\vp_n}=\frac{-\al}{8}(1-\frac{r^2}{A})^2\com
\pl^2\vp_n=\frac{8A^{-1}}{(1-A^{-1}r^2)^2}\com              \nn\\
\left.\intx\sqg\right|_{\vp_n}=\intx~e^{\vp_n}
=\frac{8\pi}{-\al}A(\frac{kA}{\ep}-1)\com                   \nn\\
\intx~\vp_n\pl^2\vp_n
=8\pi \{2-(\frac{kA}{\ep}-1)ln~(\frac{-\al}{8})
+\frac{2kA}{\ep}(ln\frac{A}{\ep}-1)\ \}\com                  \nn\\
\left.\intx\sqg R^2\right|_{\vp_n}=\intx~e^{-\vp_n}(\pl^2\vp_n)^2
=\frac{-8\pi\al}{A}(\frac{kA}{\ep}-1)\com                 \label{a.5}\\
-\left.\intx\sqg R\right|_{\vp_n}=\intx~\pl^2\vp_n
=8\pi(\frac{kA}{\ep}-1)\com                                   \nn\\
\mbox{cf.\ \ }\intx\sqg R=8\pi(1-h)\mbox{\ for oriented closed surface}
                                                           \com\nn\\
\left.R\right|_{\vp_n}=-e^{-\vp_n}\pl^2\vp_n=\frac{\al}{A}<0 \pr \nn
\end{eqnarray}
%%%%%%%%%%%%%%%%%%%%%%%%%%%%%
%*** a.6   %%%%%%%%%%%%%%%%
\begin{eqnarray}
\frac{1}{\pi A}\intx\frac{1}{(1-A^{-1}r^2)^2}
=2A^{-1}\int_0^{\infty}\frac{r~dr}{(1-A^{-1}r^2)}              \nn\\
=2A^{-1}\lim_{\ep\rightarrow +0}(\int_{0\leq r^2 <A-\ep}+\int_{r^2>A+\ep})
\frac{r~dr}{(1-A^{-1}r^2)}=(\frac{kA}{\ep}-1)\com          \label{a.6}\\
\int\frac{ln~(1-x)}{(x-1)^2}=-\frac{ln~(1-x)}{x-1}-\frac{1}{x-1} \com
\int\frac{ln~(x-1)}{(x-1)^2}=-\frac{ln~(x-1)}{x-1}-\frac{1}{x-1} \pr\nn
\end{eqnarray}

%%%%%%%%%%%%%%%%%%%%%%%%%%%%%
\newpage
\begin{flushleft}
{\bf Figure Captions}
\end{flushleft}
\begin{itemize}
\item
Fig.1\q The contour of $\la$-integral in the complex $\la$-plane
\item
Fig.2\q Infra-red cut-off $L$\ in
the flat coordinates and the sphere manifold.
For simplicity, the picture is for $\al=8.$\ For general $\al$,
$(x,y,r,\sqrt{A},L)$\ is substituted by
$\sqrt{8/\al} \times (x,y,r,\sqrt{A},L)$.
\item
Fig.3\q $A\times$ Curvature $=\al(w)$\ ,\
Positive and Negative Curv. Sols., $-$Branch
\item
Fig.4a\q Log-Log Plot of $A<\intx\sqg R^2>|_c$, $w>0$,
Positive and Negative Curv. Sols., $-$Branch
\item
Fig.4b\q Linear Plot of $A<\intx\sqg R^2>|_c$,
Positive and Negative Curv. Sols., $-$Branch
\item
Fig.5\q$\ga A\times$(String Tension)$=\ga\la(w)A$,
Positive and Negative Curv. Sols., $-$Branch
\item
Fig.6\q$\ga \times$(Total Free Energy)$=-\ga\Ga^{eff}$,
Positive and Negative Curv. Sols., $-$Branch,
The $w$-independent terms ($C(A),\Ctil_2(A)$)
are omitted.
\item
Fig.7 \q $\frac{1}{A}\times$Area$=\frac{8\pi}{\al(w)}$,
Positive and Negative Curv. Sols., $-$Branch
\item
Fig.8\ Reguralization of singularity at $r=\sqrt{A}$ \ of (\ref{4.2}).
\item
Fig.9\ Punctured sphere absorbing infrared divergence
\item
Fig.10\ $\la$-integral path for $-$branch solution
\item
Fig.11\ Semiclassical result versus KPZ formula for string susceptibility
\item
Fig.12\ Temperature $T=T(w)$, Pos. Curv. Sol.
\item
Fig.13\ Entropy per unit mol, $S_{ent}(w)/(4\pi/\ga)$, Pos.Curv.Sol.
\item
Fig.14\ Schematic image of surface in each phase
\end{itemize}
%%%%%%%

\begin{thebibliography}{99}
\bibitem{DJ} %*DJ*
 E.D'Hoker and R.Jackiw,{\PR {\bf D26},3517(1982)}
\bibitem{SEI} %*SEI*
N.Seiberg,{Prog.Theor.Phys.Suppl.{\bf 102},319(1990)}
\bibitem{AJT} %*AJT*
J.Ambj{\o}rn,S.Jain and G.Thorleifsson,{\PL {\bf 307B},34(1993)}
\bibitem{AT} %*AT*
J.Ambj{\o}rn and G.Thorleifsson,{\PL{\bf 323B},7(1994)}
\bibitem{ITY1} %*ITY1*
S.Ichinose,N.Tsuda and T.Yukawa,{Preprint of Univ.of Shizuoka,
US-94-03,hep-th/9502101,to be published in Nucl.Phys.B,"Classical Solution
of Two Dimensional $R^2$-Gravity and Cross-Over Phenomenon"}
\bibitem{ITY2} %*ITY2*
S.Ichinose,N.Tsuda and T.Yukawa,{Preprint of Univ.of Shizuoka,US-95-05,
"MINBU Dstribution of Two Dimensional Quantum Gravity:~
Simulation Result and Semiclassical Analysis"}
\bibitem{KPZ}  %*KPZ*
V.G.Knizhnik,A.M.Polyakov and A.B.Zamolodchikov,{\MPL
{\bf A3},819(1988)}
\bibitem{P}  %*P*
  A.M.Polyakov,{Phys.Lett.{\bf 103B},207(1981)}
\bibitem{N2a} %*N2a*
{
The sign for the action is different from the usual convention as seen in
(\ref{3.3}).
}
\bibitem{N2b} %*N2b*
{
The uniqueness of this term, among all possible total derivatives, is shown
in Discussions(sect.6) of \cite{ITY1}.
}
\bibitem{N2c} %*N2c*
{
This is for the comparison with the 'classical limit' $c_m\ra -\infty$.
We can do the same analysis for $\ga<0$\ without any difficulty.
}
\bibitem{N2d} %*N2d*
{
In this section only,we explicitly write $\hbar$\ (Planck constant) in order
to show the perturbation structure clearly.
}
\bibitem{N2e} %*N2e*
{
The expansion parameter is the Planck constant $\hbar$\ . It is known,
in the field theory, the expansion with repect to $\hbar$\ is equivalent
to that with respect to the number of loops in the Feynman diagrams
(loop-expansion).
}
\bibitem{N2f} %*N2f*
{
This method is so popular that its naming is diverse depending on the
applied circumstances: \ mean-field method, WKB-approximation
,stationary phase approximation, effective action method,
background-field method,etc.
          }.
\bibitem{S1} %*S1*
S.Ichinose,{Preprint of SLAC, SLAC-PUB-95-6774,
to be published in \NP B,
"Renormalization of Two Dimensional $R^2$-Gravity"}
\bibitem{OV}  %*OV*
E.Onofri and M.Virasoro,{preprint TH3233}
\bibitem{Z}   %*Z*
A.B.Zamolodchikov,{Phys.Lett.{\bf 117B},87(1982)}
\bibitem{N3a} %*N3a*
{
This condition turns out to be satisfied in the following solution
          }
\bibitem{N3b} %*N3b*
{Because of the term $-\la (\int d^2x~e^{\vp}-A)$\ in $Y[A,\la]$,(\ref{3.4f}),
we see $\la$~can be interpreted as the string (surface) tension.
}
\bibitem{N3c} %*N3c*
{
Adding the effect of the outer region does not change the essential part
of the following content. See App.A for detail.
}
\bibitem{N3cc} %*N3cc*
{
In App.A, the quantities (\ref{4.3}) are evaluated using two ways of
regularization: 1) inner region and  2) inner and outer regions.
}
\bibitem{N3d} %*N3d*
{
The infrared regularization parts, $C(A)$~ and $\Ctil_2(A)$~, does not
obey the reflection rule. Those parts do not depend on $w$~ and
directly come from
the topology of the manifold. The toplogy of a sphere and that of
the infrared-regularized sphere (Fig.9) is quite different.
}
\bibitem{N4a} %*N4a*
{
Note that these vacua are meaningful only locally.
The global constraint $\intx\sqg R=8\pi$\ is violated in these
vacua and
is irrelevent for the dynamics.
}
\bibitem{N6a} %*N6a*
{
In the  text, we understand $\xi\rightarrow +1-0$.
Practically all numerical evaluations, except Fig.12 and Fig.13,
are done at $\xi=0.99$.
}
\bibitem{KN}  %*KN*
H.Kawai and R.Nakayama,{Phys.Lett.{\bf 306B},224(1993)}
\end{thebibliography}
\end{document}